\documentclass[preprint,aps,shownopacs,preprintnumbers,amsmath,amssymb,nofootinbib,epsf,epsfig]
{revtex4}

\usepackage{amstext,amssymb}
\usepackage{amsmath}
\usepackage{graphicx}
\usepackage[hyperfootnotes=true]{hyperref}
\usepackage{color}
\usepackage{array,multirow}
\usepackage{slashed} 
\usepackage{multirow}
\newcommand{\be}{\begin{equation}}
\newcommand{\ee}{\end{equation}}
\newcommand{\bea}{\begin{eqnarray}}
\newcommand{\eea}{\end{eqnarray}}
\newcommand{\nn}{\nonumber}

\def\s1{\hat s}
\def\para{\parallel}

\def\U1mt{U(1)_{L_\mu-L_\tau}}

\def\ol{\overline}
\def\nl{\nonumber\\}
\usepackage{comment}

\usepackage{slashed}
\global\long\def\d{\partial}

\begin{document}
\title{\bf Light dark matter, rare $B$ decays with missing energy in  $L_{\mu}-L_{\tau}$ model with a scalar leptoquark }
\author{Shivaramakrishna Singirala$^a$}
\email{krishnas542@gmail.com}
\author{Suchismita Sahoo$^b$}
\email{suchismita8792@gmail.com}
\author{Rukmani Mohanta$^a$}
\email{rmsp@uohyd.ac.in}
\affiliation{$^a$School of Physics,  University of Hyderabad, Hyderabad-500046,  India\\
$^b$Department of Physics, Central University of Karnataka, Kalaburagi-585367, India}


\begin{abstract}
We investigate the phenomenology of light GeV-scale fermionic dark matter in $U(1)_{L_\mu - L_{\tau}}$ gauge extension of the Standard Model. Heavy neutral fermions alongside with a $S_1(\overline{3}$,$1$,$1/3$) scalar leptoquark and an inert scalar doublet are added to address the flavor anomalies and light neutrino mass respectively. The light gauge boson associated with $U(1)_{L_\mu-L_\tau}$ gauge group mediates dark to visible sector and helps to obtain the correct relic density. Aided with a colored scalar, we constrain the new model parameters by using the branching ratios of various $b \to sll$ and $b \to s \gamma$ decay processes as well as the lepton flavour non-universality observables $R_{K^{(*)}}$ and then show the implication on the branching ratios of  some rare semileptonic  $B \to (K^{(*)}, \phi)+$ missing energy, processes.  


\end{abstract}

\maketitle
\flushbottom
\section{Introduction} 

Standard Model (SM) of particle physics  is quite successful in meeting the experimental sensitivities when it comes to interactions at fundamental level. It is believed to be a low-energy gauge theory embedded in a high scale unified version. Despite its accomplishments, the increasing sensitivities of various intensity and cosmic frontier experiments have spotlighted its drawbacks and hints towards its extension. Listing a few, the existence of dark matter (DM) and its nature \cite{Zwicky:1937zza, Rubin:1970zza, Clowe:2003tk,Bertone:2004pz,ArkaniHamed:2008qn,Dodelson:1993je}, non-zero neutrino masses and its mixing phenomena \cite{Zyla:2020zbs}, matter-antimatter asymmetry \cite{Sakharov:1967dj,Kolb:1979qa, Davidson:2008bu,Buchmuller:2004nz,Strumia:2006qk} etc.  

Though most of the flavor observables go along with the SM, there are a collection of recent measurements in  semileptonic $B$ meson decays, involving $b \to sll$ and $b \to c \tau \bar \nu_\tau$ quark level transitions, that are incongruous with the SM predictions. The most conspicuous measurements, hinting the physics beyond SM are the lepton flavor universality violating parametes:  $R_{K}$ with a discrepancy of $3.1\sigma$ \cite{Aaij:2014ora, Aaij:2019wad, Aaij:2021vac, Bobeth:2007dw, Bordone:2016gaq}, $R_{K^{(*)}}$ with a disagreement at the level of $(2.1-2.5)\sigma$ \cite{Aaij:2017vbb, Capdevila:2017bsm}, $R_{D^{(*)}}$ with $3.08\sigma$ discrepancy \cite{Amhis:2019ckw, Na:2015kha, Fajfer:2012vx, Fajfer:2012jt} and $R_{J/\psi}$ with a deviation of nearly $2\sigma$ \cite{Aaij:2017tyk, Wen-Fei:2013uea, Ivanov:2005fd} from their SM predictions. Though the Belle Collaboration \cite{Abdesselam:2019lab, Abdesselam:2019wac} has also announced their measurements on $R_{K^{(*)}}$ in various $q^2$ bins, however these measurements have large uncertainties.  Besides the $R_{K^{(*)}}$ parameters,  the $P_5^\prime$ optimized observable disagrees with the SM at the level of $4\sigma$ in the $(4.3-8.68)~\rm {GeV}^2$ $q^2$-bin \cite{Aaij:2013qta, Aaij:2015oid, Abdesselam:2016llu} and the decay rate  of  $B \to K^* \mu \mu$ shows $3\sigma$ discrepancy \cite{Aaij:2014pli}. The branching ratio of $B_s \to \phi \mu \mu$ channel also disagrees with the theory at the level of $3\sigma$ \cite{Aaij:2013aln} in low $q^2$.

Due to the above discussed anomalies in $b \to sll$, the rare semileptonic $B$ decays with charged leptons in the final state such as $B \to K^{(*)} ll$, $B_s \to \phi ll$  have attracted large attention in recent times compared to the analogous semileptonic $B$ meson channels with neutral leptons in the final state, i.e., $B \to K^{(*)} \nu \bar \nu$, $B_s \to \phi \nu \bar \nu$. Since the neutrinos escape undetected, the number of angular observables associated with charged leptons are also more in contrast to the $b \to s \nu \bar \nu$ processes.  In the SM, the rare  $b \to s \nu \bar \nu$ transitions are significantly suppressed by the loop momentum and off-diagonal CKM matrix elements. Even though these processes are theoretically very clean compared to other FCNC decays, these are yet to be observed; there exist only  upper limits in the branching ratios of such decays \cite{Zyla:2020zbs}.   The theoretical computation of inclusive $B \to X_s \nu \bar \nu$ process is quite easy, however the experimental measurement of this decay is probably unfeasible due to the requirement of reconstruction of all $X_s$ and  the missing neutrinos. However, the exclusive $B \to K^{(*)} \nu \bar \nu$ processes are more promising  to look for  physics beyond the SM.   

Of particular interest among the FCNC $B$  decays are the  semileptonic decays of the form $b \to s +\displaystyle{\not}{E}$, where $\displaystyle{\not}E$ stands for the missing energy. Besides the production of heavy meson in charge and CP correlated states, the existing $B$-factories can tag the missing energy decays of  $B$ meson ``on the other side". In  Ref. \cite{Badin:2010uh},  the leptonic decays of heavy $B(D)$ meson to a pair of  scalar or fermion or vector  particles has been investigated both  in model independent way and in some popular models.  Many new physics models \cite{Aslam:2008th, Aliev:2007gr, Altmannshofer:2009ma, Kim:2007uq, Jeon:2006nq, Kim:2009mp, Sirvanli:2007yq, Smith:2010st, Mahajan:2003gx} also have  been proposed to test the observed rates of  $B \to M + \displaystyle{\not} E$  decays.  Furthermore, the sensitivity of bounds on WIMP-nucleon cross section for GeV scale dark matter in direct detection experiments is less. This brings us to study a light GeV scale WIMP ($< 2.5$ GeV) in the light of missing energy in rare $B$-decays. 

To input our purpose, we opt for a simple but phenomenologically rich the $U(1)_{L_\mu - L_\tau}$ gauge extension. Proposed by X.G. He et al, in \cite{He:1991qd,He:1990pn}, and then later explored to explain muon $g-2$, dark matter, neutrino phenomenology and matter-antimatter asymmetry of the Universe \cite{Ma:2001md,Baek:2008nz,Altmannshofer:2014cfa,Heeck:2014qea,Crivellin:2015mga,Fuyuto:2015gmk,Patra:2016ofq,Biswas:2016yan,Altmannshofer:2016jzy,Araki:2017wyg,Chen:2017cic,Chen:2017usq,Baek:2017sew,Bauer:2018onh,Kamada:2018zxi,Gninenko:2018tlp,Nomura:2018vfz,Banerjee:2018mnw,Heeck:2018nzc,Escudero:2019gzq,Altmannshofer:2019zhy,Biswas:2019twf,Kowalska:2019ley,Kang:2019vng,Joshipura:2019qxz,Han:2019diw,Jho:2020jsa,Amaral:2020tga,Borah:2020jzi,Huang:2021nkl,Borah:2021jzu}. In our previous work \cite{Singirala:2018mio}, we extended this model with three heavy neutral fermions, a $(\overline{3},1,1/3)$ scalar leptoquark and investigated heavy fermion DM and various flavor anomalies. In the present  work, we look at low mass regime of DM arising as a missing energy in semileptonic decays of  $B$ meson  to $M(=K, K^*, \phi)$. 
The extant of  LQ is proposed  in many  theoretical frameworks, such as the grand unified theories \cite{Georgi:1974sy, Fritzsch:1974nn, Langacker:1980js, Georgi:1974my}, Pati-Salam model \cite{Pati:1974yy, Pati:1973uk, Pati:1973rp, Shanker:1981mj, Shanker:1982nd}, quark and lepton composite model \cite{Kaplan:1991dc} and the technicolor model \cite{Schrempp:1984nj, Gripaios:2009dq}.  The study of flavor anomalies as well as $B$ anomalies in connection with the dark matter with scalar LQ have been investigated in the literature \cite{Alok:2017sui, Becirevic:2017jtw, Hiller:2017bzc, DAmico:2017mtc, Becirevic:2016yqi, Bauer:2015knc, Bordone:2016gaq, Calibbi:2015kma, Freytsis:2015qca, Dumont:2016xpj, Dorsner:2016wpm, Varzielas:2015iva, Dorsner:2011ai,  Davidson:1993qk,  Saha:2010vw,  Mohanta:2013lsa, Sahoo:2015fla, Sahoo:2015pzk, Sahoo:2015qha, Sahoo:2015wya, Kosnik:2012dj, Singirala:2018mio, Chauhan:2017ndd, Becirevic:2018afm, Angelescu:2018tyl, Sahoo:2017lzi, Sahoo:2016nvx, Sahoo:2016edx, Sahoo:2020wnk}.

The paper is organized as follows. In section-II, we provide the details of model, mixing in gauge, fermion and scalar sectors. Invisible widths of Higgs and $Z$ boson are given in section-III. We discuss light dark matter relic density and its detection prospects in section-IV. Comments on light neutrino mass is given in section-V. Section-VI contains the detailed discussion on further  constraints on new parameters from the flavor anomalies. The study of $B_{(s)} \to (K^{(*)}, \phi) +\displaystyle{\not} E$ is presented in section-VII.  Finally the conclusive remarks are provided in section-VIII.
\section{Model Description}
We consider a variant of $L_{\mu}-L_{\tau}$ model \cite{Singirala:2018mio}, recently explored in the context of flavor anomalies and dark matter with three additional neutral fermions $N_{e}, N_{\mu}, N_{\tau}$ and a $S_{1}(\bar{3},1,1/3)$ scalar leptoquark (SLQ). Alongside, the model comprises of an additional inert scalar doublet $\eta$  for generating neutrino mass at one-loop and a singlet $\phi_2$ that spontaneously breaks the new $U(1)$ symmetry. The full list of field content is provided in Table.  \ref{lmutau_model}.
\begin{table}[htb]
\begin{center}
\begin{tabular}{|c|c|c|c|c|}
	\hline
			& Field	& $ SU(3)_C \times SU(2)_L\times U(1)_Y$	& $U(1)_{L_{\mu}-L_{\tau}}$	& $Z_2$\\
	\hline
	\hline
	Fermions	& $Q_L \equiv(u, d)^T_L$			& $(\textbf{3},\textbf{2}, 1/6)$	& $0$	& $+$\\
			& $u_R$							& $(\textbf{3},\textbf{1}, 2/3)$	& $0$ & $+$	\\
			& $d_R$							& $(\textbf{3},\textbf{1},-1/3)$	& $0$	& $+$\\
			& $\ell_L \equiv e_L,\mu_L,\tau_L$	& $(\textbf{1},\textbf{2},  -1/2)$	&  $0,1,-1$	& $+$\\
			& $\ell_R \equiv e_R,\mu_R,\tau_R$							& $(\textbf{1},\textbf{1},  -1)$	&  $0,1,-1$	& $+$\\
			& $N_{e},N_{\mu},N_{\tau}$						& $(\textbf{1},\textbf{1}, 0)$	&  $0,1,-1$	& $-$\\
	\hline
	Scalars	& $H$							& $(\textbf{1},\textbf{2},~ 1/2)$	&   $0$	& $+$\\
		& $\eta$							& $(\textbf{1},\textbf{2},~ 1/2)$	&   $0$	& $-$\\
			& $\phi_2$						& $(\textbf{1},\textbf{1},~   0)$	&  $2$	& $+$\\  
			& $S_1$						& $(\bar{\textbf{3}},\textbf{1},   1/3)$	&  $-1$	& $-$\\    
	\hline
	Gauge bosons			& $W_\mu^i ~(i=1,2,3)$							& $(\textbf{1},\textbf{3},0)$	&   $0$	& $+$\\
				& $B_\mu$							& $(\textbf{1},\textbf{1},0)$	&   $0$	& $+$\\
						& $V_\mu$							& $(\textbf{1},\textbf{1},0)$	&   $0$	& $+$\\
	\hline
	\hline
\end{tabular}
\caption{Fields in the chosen $U(1)_{L_{\mu}-L_{\tau}}$ model.}
\label{lmutau_model}
\end{center}
\end{table}

The relevant Lagrangian terms corresponding to gauge, fermion, gauge-fermion interaction and scalar sectors  are given by
\begin{eqnarray}
&&\mathcal{L}_{G} = -{1 \over 4} \left(\bold{\hat W}_{\mu\nu}\bold{\hat W}^{\mu\nu} + {\hat B}_{\mu\nu} {\hat B}^{\mu\nu} + {\hat V}_{\mu\nu} {\hat V}^{\mu\nu} + 2 \sin\chi {\hat B}_{\mu\nu} {\hat V}^{\mu\nu}\right),\nl
&&\mathcal{L}_{f} = -\frac{1}{2}M_{ee}\ol{N_{e}^c} N_{e} -\frac{f_\mu}{2}\left({\ol{N_{\mu}^c}} N_{\mu}\phi_2^{\dagger}+{\rm h.c.}\right) - \frac{f_\tau}{2}\left({\ol{N_{\tau}^c}} N_{\tau}\phi_2 + {\rm h.c.} \right) -\frac{1}{2}M_{\mu\tau}(\ol{N_{\mu}^c} N_{\tau} + \ol{N_{\tau}^c} N_{\mu}), \nl 
&& ~~~~~~~-\sum_{l=e,\mu,\tau} (Y_{l l} (\ol{\ell_L})_l \tilde \eta N_{lR} + {\rm{h.c}})- \sum_{q=d,s,b} (y_{q R}\; \ol{d_{qR}^c} S_1 N_{\mu} + {\rm{h.c.}}),\nl
&&\mathcal{L}_{G-f} = -g_{\mu\tau} \overline{\mu} \, \gamma^\mu \mu {\hat V}_\mu  +g_{\mu\tau}\overline{\tau}  \gamma^\mu \tau {\hat V}_\mu - g_{\mu\tau} \overline{\nu_{\mu}} \, \gamma^\mu (1-\gamma^5) \nu_{\mu} {\hat V}_\mu + g_{\mu\tau} \overline{\nu_{\tau}} \, \gamma^\mu (1-\gamma^5) \nu_{\tau} {\hat V}_\mu \nn\\
&&~~~~~~~~~- g_{\mu\tau}\overline{N_\mu}{\hat V}_\mu \gamma^\mu \gamma^5 N_{\mu} + g_{\mu\tau}\overline{N_\tau} {\hat V}_\mu \gamma^\mu \gamma^5 N_{\tau},\nl
&&\mathcal{L}_S = \left|\left(i \d_\mu - \frac{g}{2} \boldsymbol{\tau}^a\cdot\bold{\hat W}_\mu^a  -\frac{g^{\prime}}{2}{\hat B}_\mu\right) \eta \right|^2 
+ \left| \left(i \d_\mu -\frac{g^{\prime}}{3}{\hat B}_\mu + g_{\mu\tau} \,{\hat V}_\mu  \right) S_1\right|^2 + \left| \left(i \d_\mu -2 g_{\mu\tau} \,{\hat V}_\mu  \right) \phi_2\right|^2 \nl
&&~~~~- V(H,\eta,\phi_2,S_1),
\label{Lag}
\end{eqnarray}
where, the scalar potential is expressed as
\begin{align}
V(H,\eta,\phi_2,S_1) &=  \mu^2_{H}(H^{\dagger}H)+ \lambda_{H}(H^{\dagger}H)^2 + \mu^2_{\eta}(\eta^{\dagger}\eta)+ \lambda_{H\eta}(H^{\dagger}H)(\eta^{\dagger}\eta) + \lambda_{\eta}(\eta^{\dagger}\eta)^2 + \lambda'_{H\eta}(H^{\dagger}\eta)(\eta^{\dagger}H) \nn\\
      +& \frac{\lambda''_{H\eta}}{2}\left[(H^{\dagger}\eta)^2 + {\rm h.c.}\right] 
 + \mu^2_{\phi} (\phi^\dagger_2 \phi_2) + \lambda_{\phi} (\phi^\dagger_2 \phi_2)^2 
      +\mu^2_{S} ({S_1}^\dagger {S_1})  +\lambda_{S} ({S_1}^\dagger {S_1})^2 \nonumber \\
      +&  \left[\lambda_{H\phi} (\phi^\dagger_2 \phi_2) + \lambda_{HS} (S^\dagger_1 S_1)\right](H^\dagger H)+  \lambda_{S\phi}(\phi^\dagger_2 \phi_2) ({S_1}^\dagger {S_1}) +\lambda_{\eta \phi}(\phi^\dagger_2 \phi_2) (\eta^\dagger \eta) \nn\\
      +& \lambda_{S\eta}({S_1}^\dagger S_1) (\eta^\dagger \eta).
\label{eq:potential}
\end{align}
In the above, $\mu^2_\phi, \mu^2_H < 0$, the scalar $\phi_2$ breaks $U(1)_{L_\mu-L_\tau}$  spontaneously with $\langle \phi_2 \rangle = v_2/\sqrt{2}$ and $SU(2)_L\times U(1)_Y$ gets spontaneously broken by $H$ with $\langle H \rangle = v/\sqrt{2}$. The minimisation conditions are as follows,
\begin{eqnarray}
\mu_H^2 &=& -\lambda_H v^2 - \lambda_{H\phi} v_2^2/2, \nn\\
\mu_\phi^2 &=& -\lambda_\phi v_2^2 - \lambda_{H\phi} v^2/2.\nn
\end{eqnarray}
We have $\mu^2_\eta, \mu^2_S > 0$ and the masses of the colored scalar and inert components (charged and neutral) are
\begin{eqnarray}
M_{S_1}^2 &=& 2\mu_S^2 + \lambda_{HS}v^2 + \lambda_{S\phi}v_2^2\;,\nn\\
M^2_{\eta_c} &=& \mu_\eta^2 + \lambda_{H\eta} v^2/2 + \lambda_{\eta\phi} v_2^2/2,\nn\\
M^2_{\eta_{r,i}} &=& \mu_\eta^2 + \left(\lambda_{H\eta} + \lambda^\prime_{H\eta} \pm \lambda^{\prime\prime}_{H\eta}\right) v^2/2 + \lambda_{\eta\phi} v_2^2/2.
\end{eqnarray}
In our analysis, we consider the breaking scale of new $U(1)$ to be above the electroweak scale. The CP-odd component of $\phi_2$ gets absorbed by the new $U(1)$ associated gauge boson, which plays a major role in the subsequent phenomenological study. The mass of colored scalar is taken to be $1.2$ TeV \footnote{The most relevant process for the SLQ search at LHC is the pair production through $gg (q \bar q) \to S_1^\dagger S_1$. Both ATLAS and CMS have searched for this production process through different LQ decay channels into second and third generation quarks and leptons:  $S_1^\dagger S_1\to t \bar{t} l \bar{l}, b \bar{b} \nu \bar{ \nu}$. As a consequence, these searches provide  suitable model-independent constraints both on mass and branching fractions of the LQ. The best limits on the masses of second/third generation LQs obtained for benchmark branching ratio values set to $\beta^\prime=1(0.5)$ are as follows: $900 (560)$ GeV from $t \bar{ t} \tau \bar{ \tau}$ and  $1100 (800)$ GeV from $b \bar{ b} \nu \bar{ \nu}$ channels \cite{Faroughy:2018cyl}.}
. 
\subsection{Gauge mixing}
For the mixing of the gauge bosons of $U(1)_Y$ and $U(1)_{L_\mu-L_\tau}$ symmetries, we use the transformation \cite{Napsuciale:2020yvu}
\begin{align}
	\begin{pmatrix}
		 \bar{B}_\mu	\\
		 \bar{V}_{\mu}	\\
	\end{pmatrix}
    	=
	\begin{pmatrix}
		 ~1~   & ~\sin\chi~  \\
 ~0~  & ~\cos\chi~	\\
	\end{pmatrix}\begin{pmatrix}
		 \hat{B}_\mu	\\
		 \hat{V}_{\mu}	\\
	\end{pmatrix}.
\end{align}
Thus, the gauge kinetic terms take canonical form
\begin{eqnarray}
\mathcal{L'}_G &=& -{1 \over 4} \left(\bold{\hat W}_{\mu\nu}\bold{\hat W}^{\mu\nu} + {\bar B}_{\mu\nu} {\bar B}^{\mu\nu} + {\bar V}_{\mu\nu} {\bar V}^{\mu\nu} \right).
\end{eqnarray}
Expanding the scalar covariant derivatives, we obtain the mass matrix of gauge fields in the basis $\left(W^3_{\mu},\bar{B}_\mu,\bar{V}_\mu\right)$ as
\begin{align}
M^2_{G}	=
	\begin{pmatrix}
		 ~\frac{1}{8}g^2 v^2~   & ~-\frac{1}{8}gg^\prime v^2~ & ~\frac{1}{8}gg^\prime \tan\chi v^2~  \\
 ~-\frac{1}{8}gg^\prime v^2~  & ~\frac{1}{8}g^{\prime 2} v^2~ & ~-\frac{1}{8}g^{\prime 2} \tan\chi v^2~	\\
~\frac{1}{8}gg^\prime \tan\chi v^2~ & ~-\frac{1}{8}g^{\prime 2} \tan\chi v^2 & ~2g^2_{\mu\tau} \sec\chi^2 v^2~
	\end{pmatrix}.
	\label{matrix_gauge}
\end{align}
Two unitary matrices are required to get to the mass eigenstate basis of gauge bosons, given as follows. 
\begin{equation}
U_1^T M^2_G ~U_1 = 	\frac{1}{2}\begin{pmatrix}
		 M^{2}_{Z_{SM}}   & ~0~ & ~\delta M^2~  \\
 ~0~  & ~0~ & ~0~	\\
~\delta M^2 & ~0~ & ~M_{\bar{V}}^2~
	\end{pmatrix} ~{\rm with}~ U_1
	=
	\begin{pmatrix}
		 \cos{\theta_w}	& \sin{\theta_w} & 0	\\
		 -\sin{\theta_w}	& \cos{\theta_w} & 0	\\
		 0 & 0 & 1
	\end{pmatrix}. 
\end{equation}
In the above,
\begin{eqnarray}
&& M^2_{Z_{SM}}= \frac{1}{4} (g^2 + g^{\prime 2})v^2, \nn\\
&& M^2_{\bar{V}} = 4 g_{\mu\tau}^2 \sec \chi^2 v_2^2, \nn\\
&& \delta M^2 = \frac{1}{4} g^\prime \sqrt{(g^2 + g^{\prime 2})} \tan\chi v^2, \nn\\
&& \theta_w = \tan^{-1} [g^\prime/g]
\end{eqnarray}
Further operating with $U_2$, we obtain
\begin{align}
(U_1 U_2)^T M^2_{G} ~(U_1 U_2) =
\frac{1}{2}\begin{pmatrix}
		 M^{2}_{Z}   & ~0~ & ~0~  \\
 ~0~  & ~0~ & ~0~	\\
~0~ & ~0~ & ~M_{Z^\prime}^2
	\end{pmatrix} ~{\rm with}~ U_2
	=
	\begin{pmatrix}
		 \cos{\alpha}	& 0 &  \sin{\alpha} 	\\
		 0 & 1 & 0 \\
		 -\sin{\alpha}	& 0 & \cos{\alpha} 	\\
	\end{pmatrix},
\end{align}
The masses of physical gauge fields and corresponding mixing angle read as 
\begin{eqnarray}
&& M^2_{Z} = M^2_{Z_{SM}} \cos\alpha^2 - \delta M^2 \sin 2\alpha + M^2_{\bar{V}}\sin\alpha^2,\nn\\
&& M^2_{Z^\prime} = M^2_{Z_{SM}} \sin\alpha^2 + \delta M^2 \sin 2\alpha + M^2_{\bar{V}}\cos\alpha^2,\nn\\
&&\alpha = \frac{1}{2} \tan^{-1} \left[\frac{2~\delta M^2}{M^2_{\bar{V}}-M^{2}_{Z_{SM}}}\right].
\end{eqnarray}
Thus, the gauge and mass eigenstates can be related by
\begin{align}
	\begin{pmatrix}
		 \hat{W}^3_{\mu}	\\
		 \bar{B}_{\mu}	\\
		 \bar{V}_{\mu}	\\
	\end{pmatrix}	=
	U_1  U_2 \begin{pmatrix}
{Z}_{\mu}	\\
		 {A}_{\mu}	\\
		 {Z}_{\mu}^\prime	\\
		 \end{pmatrix}.
\end{align}
In the low mass regime of $Z^\prime$, the kinetic mixing gets severely constrained ($\chi \sim 10^{-3}$) from electroweak precision data \cite{Hook:2010tw,Cline:2014dwa}, and the mixing angle $\alpha \sim 4.84\times 10^{-4}$. 

Bounds on the new gauge parameters ($M_{Z^\prime}$ and $g_{\mu\tau}$) are levied by various collider experiments. Upper limits are provided by BABAR \cite{TheBABAR:2016rlg} from the cross section of $\sigma(e^+e^-\to\mu^+\mu^-Z^\prime, Z^\prime \to \mu^+\mu^-)$ and also from CMS \cite{Sirunyan:2018nnz} from the process $pp \to Z\mu\mu, Z \to Z^\prime \mu\mu$ going to a $4\mu$ final state. Other bounds from Belle \cite{Adachi:2019otg} are from the invisible decays of $Z^\prime$ as missing energy in $e^+ e^-$ collision. Stringent limits are provided from the study of neutrino trident production from CCFR collaboration \cite{Mishra:1991bv,Altmannshofer:2014pba} and DUNE \cite{Altmannshofer:2019zhy}. 
\subsection{Scalar and Fermion mixing}
The CP-even scalars $h$ and $h_2$ mix, so as heavy fermion states $N_\mu$ and $N_\tau$. The mixing matrices of both scalar and fermion sectors is given by
\begin{align}
M_H^2
	=
	\begin{pmatrix}
		 2 \lambda_H v^2   & {\lambda}_{H\phi} {v}v_2  \\
 {\lambda}_{H\phi} {v}v_2  & 2 \lambda_{\phi} v^2_2				\\
	\end{pmatrix} \;,
    \quad 	M_N
	=
	\begin{pmatrix}
		 \frac{1}{\sqrt{2}}f_{\mu}v_2	& M_{\mu\tau}	\\
		 M_{\mu\tau}	& \frac{1}{\sqrt{2}}f_{\tau}v_2				\\
	\end{pmatrix} \;.
\end{align}
One can diagonalize the above mass matrices by $U_{\beta(\zeta)}^T M_{N(H)} U_{\beta(\zeta)} = {\rm{diag}}~[M_{{-}(H_{1})},M_{{+}(H_{2})}]$, where $U_{\beta(\zeta)}$ is a $2\times 2$ unitary matrix.  The mixing angles read as $\zeta = \frac{1}{2}\tan^{-1}\left(\displaystyle{\frac{\lambda_{H\phi} v v_2}{\lambda_\phi v^2_2 - \lambda_H v^2}}\right)$ and $\beta = \frac{1}{2}\tan^{-1}\left(\displaystyle{\frac{2M_{\mu\tau}}{(f_\tau - f_{\mu})(v_2/\sqrt{2})}}\right)$. The couplings and mass eigenvalues are related as
\begin{eqnarray}
&& f_{\mu} = \frac{\sqrt{2}}{v_2}\left(M_{-}\cos^2\beta + M_{+}\sin^2\beta\right),\nn\\
&& f_{\tau} = \frac{\sqrt{2}}{v_2}\left(M_{-}\sin^2\beta + M_{+}\cos^2\beta\right),\nn\\
&& M_{\mu\tau} = \cos\beta\sin\beta \left(M_{+} - M_{-}\right),\nn\\
&& \lambda_{H} = \frac{1}{2 v^2}\left(M^2_{H_1}\cos^2\zeta + M^2_{H_2}\sin^2\zeta\right),\nn\\
&& \lambda_{\phi} = \frac{1}{2 v_2^2}\left(M^2_{H_1}\sin^2\zeta + M^2_{H_2}\cos^2\zeta\right),\nn\\
&& \lambda_{H\phi} = \frac{1}{v v_2}\cos\zeta\sin\zeta \left(M^2_{H_2} - M^2_{H_1}\right).
\end{eqnarray}
In case of scalar sector with minimal mixing ($\sin\zeta < 10^{-2}$), the mass eigenstate $H_1$ is assumed to be observed Higgs at LHC ($M_{H_1} \sim 125$ GeV) and $H_2$ is considered to be heavier one with mass 500 GeV. In the fermion spectrum, $N_-$ is the stable light dark matter with mass less than $2.5$ GeV and $M_{+} = 500$ GeV, with the corresponding mixing parameter $\sin\beta = 1/2$. All the mentioned values for model parameters are provided in Table. \ref{Tab:input para}, which will be utilized for the analysis in the subsequent sections.
 \begin{table}[htb]
 \centering
 \begin{tabular}{|c|c|c|c|c|c|c|c|c|c|}
 \hline
~Parameters~&~$M_{S_1}$ [GeV]~&~$M_{+}$ [GeV]~&~$M_{H_1}$ [GeV]~&~$M_{H_2}$ [GeV]~&~$\sin\beta$~&~$\sin\zeta$~&~$\chi$~&~$\alpha \times 10^{4}$\\
 \hline
 ~Values~&~$1200$~&~$500$~&~$125$~&~$500$~&~$1/2$~&~$10^{-3}-10^{-2}$~&~$10^{-3}$~&~$4.83-4.85$\\
 \hline
 \end{tabular}
 \caption{Model parameters along with their values in the present model.}\label{Tab:input para}
 \end{table}
\section{Invisible widths}
In the present work, Higgs ($H_1$) and $Z$ boson can decay to $N_-N_-$ and the corresponding expressions for invisible widths are given as follows
\begin{equation}
\Gamma^{H_1}_{\rm inv} = \frac{(f_\mu \cos^2\beta + f_\tau \sin^2\beta)^2\sin^2\zeta}{8\pi}  ~M_{H_1} \left(1- \frac{4 M_{-}^2}{M_{H_1}^2}\right)^{\frac{3}{2}}.
\end{equation}
$\Gamma^{Z}_{\rm inv} = \Gamma^Z_{\nu\bar{\nu}} + \Gamma^Z_{N_- N_-}$, where
\begin{eqnarray}
&&\Gamma^Z_{\nu\bar{\nu}} = \frac{g^2(\cos\alpha - \sin\alpha \tan\chi \sin\theta_w)^2}{96\pi \cos^2\theta_w} M_Z, \nn\\ 
&&\Gamma^Z_{N_- N_-} = \frac{(g_{\mu\tau}\cos 2\beta \sec\chi \sin\alpha)^2}{24\pi}  M_{Z} \left(1- \frac{4 M_{-}^2}{M_Z^2}\right)^\frac{3}{2}.
\end{eqnarray}
\begin{figure}[thb]
\begin{center}
\includegraphics[width=0.48\linewidth]{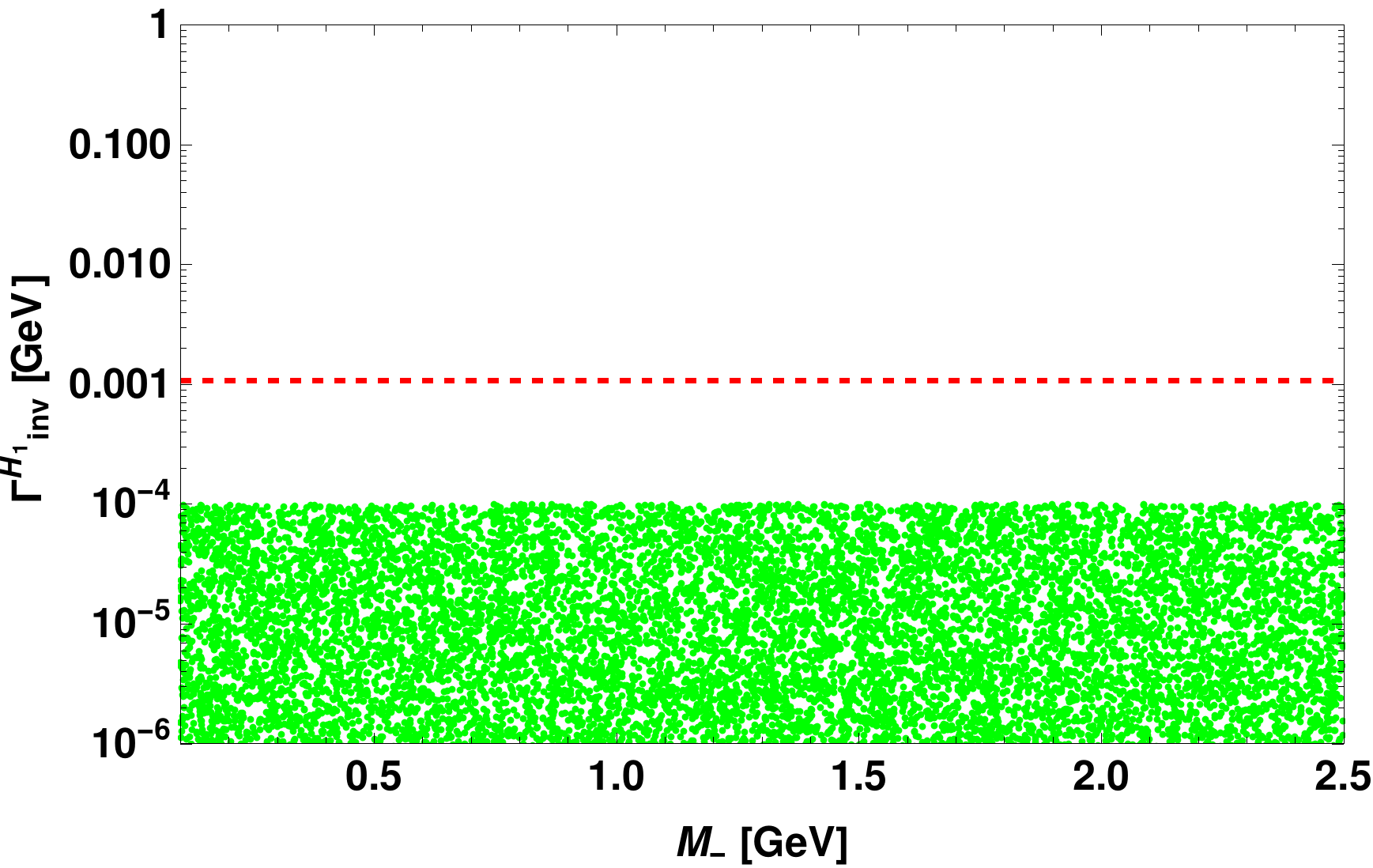}
\vspace{0.1 cm}
\includegraphics[width=0.48\linewidth]{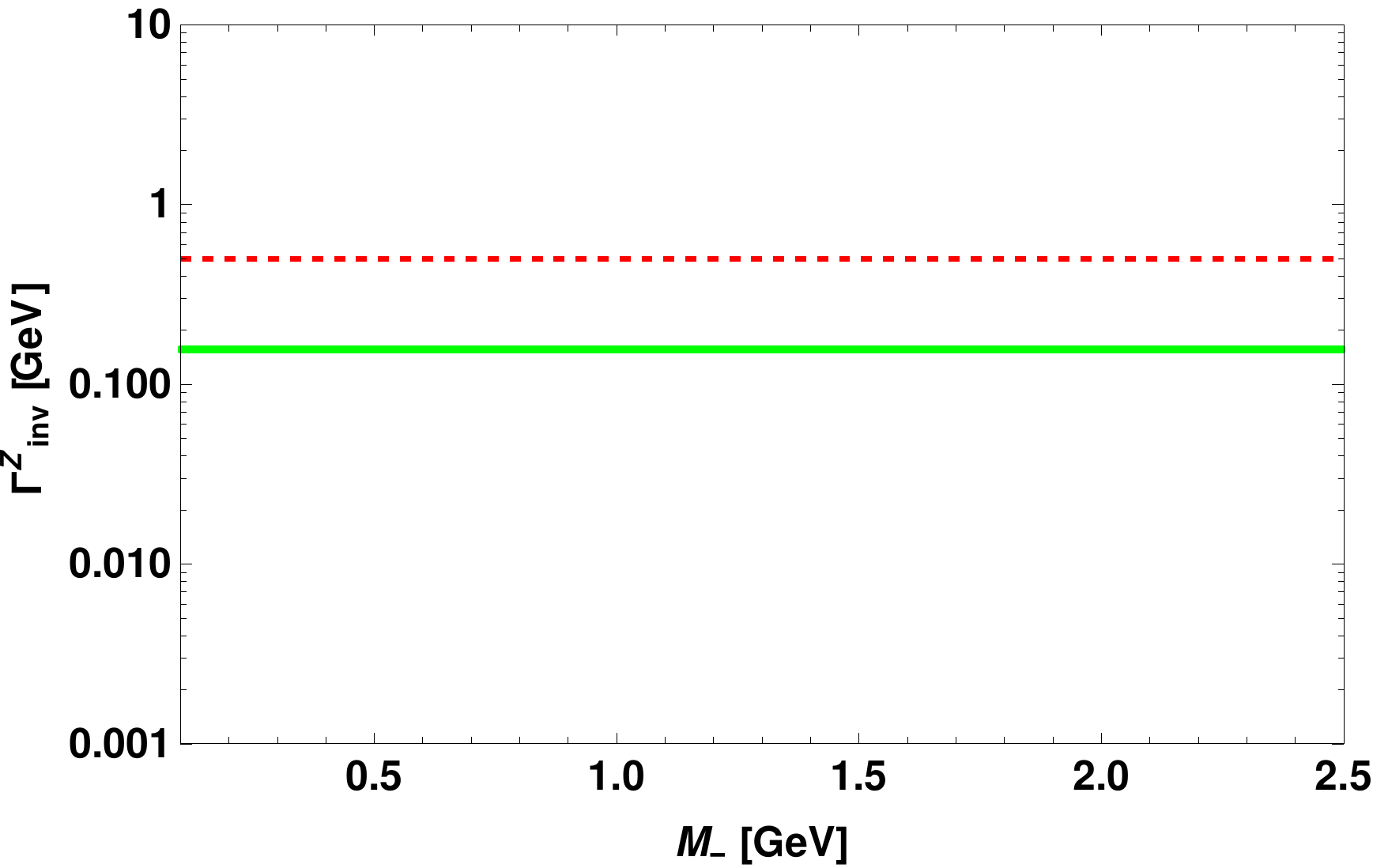}
\caption{Left panel projects invisible width of Higgs ($H_1$) for $\sin\zeta < 10^{-2}$ and right panel projects the same for $Z$ boson. Corresponding experimental upper limits \cite{Aaboud:2019rtt,ALEPH:2005ab} are denoted by red horizontal dashed lines.}
\label{relic_curve}
\end{center}
\end{figure}
%
\section{Dark matter}
\subsection{Abundance}
To compute the relic density of the light DM ($N_-$) via freeze-out mechanism, we use LanHEP \cite{Semenov:1996es} and micrOMEGAs packages \cite{Pukhov:1999gg, Belanger:2006is, Belanger:2008sj}. The channels with lepton-anti lepton pair in final state ($\mu \overline{\mu}$, $\tau \overline{\tau}$, $\nu_\mu \overline{\nu_\mu}$, $\nu_\tau \overline{\nu_\tau}$) via $Z^\prime$ and $\eta$ portal contribute to relic density. Furthermore, SLQ portal t-channel processes with $d\overline{d}$, $s\overline{s}$ in the final state are also kinematically allowed. The key point is that the s-channel processes via light $Z^\prime$ provide a resonance in propagator, thereby meeting the Planck relic density value \cite{Aghanim:2018eyx} for the DM mass in the range $0.1-2.5$ GeV. The same is visible from left panel of Fig. \ref{relic_curve}, which projects relic density of DM with its mass. Right panel shows the $3\sigma$ Planck allowed region in $M_{Z^\prime}-M_-$ plane. 

%
\begin{figure}[thb]
\begin{center}
\includegraphics[width=0.48\linewidth]{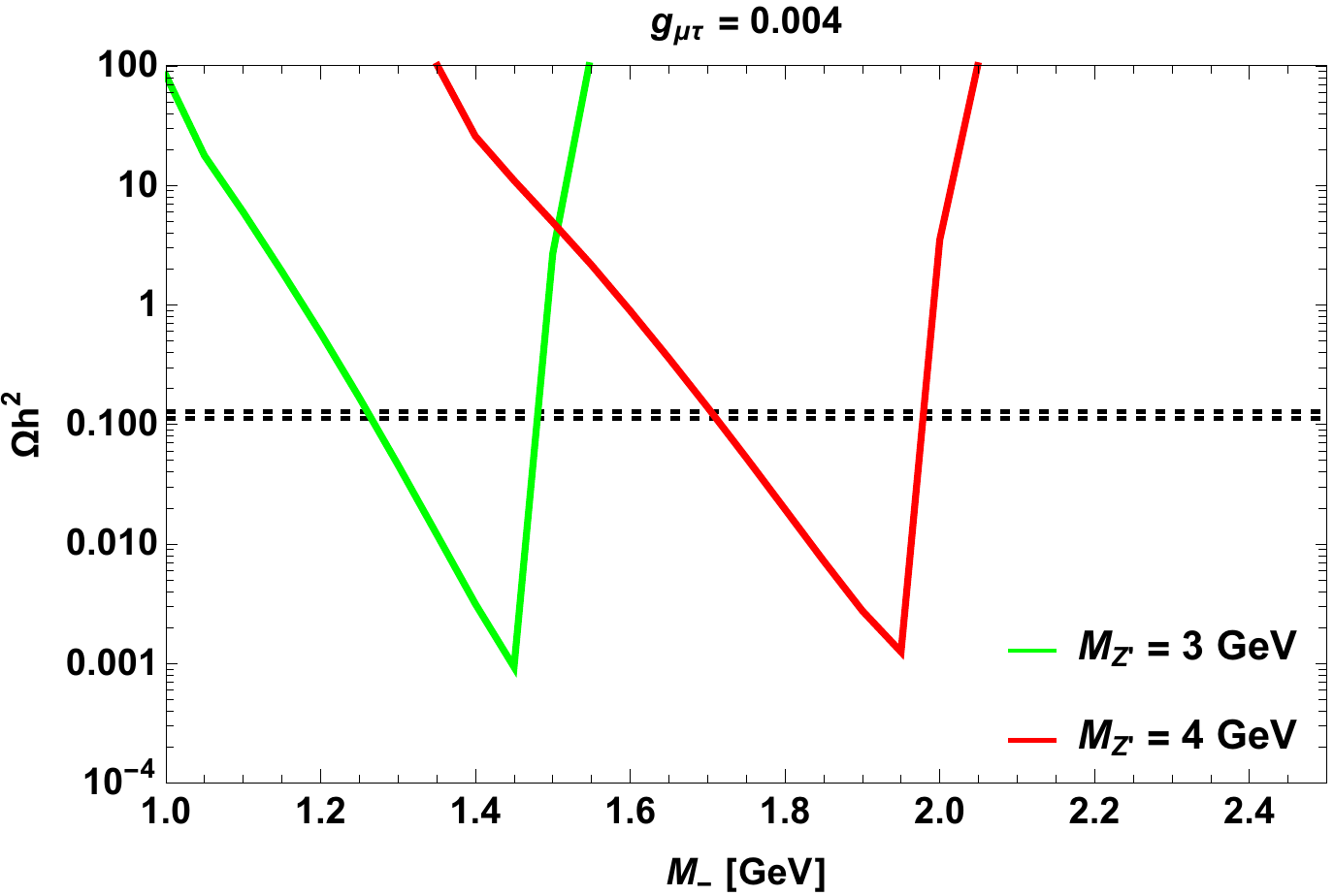}
\vspace{0.1 cm}
\includegraphics[width=0.48\linewidth]{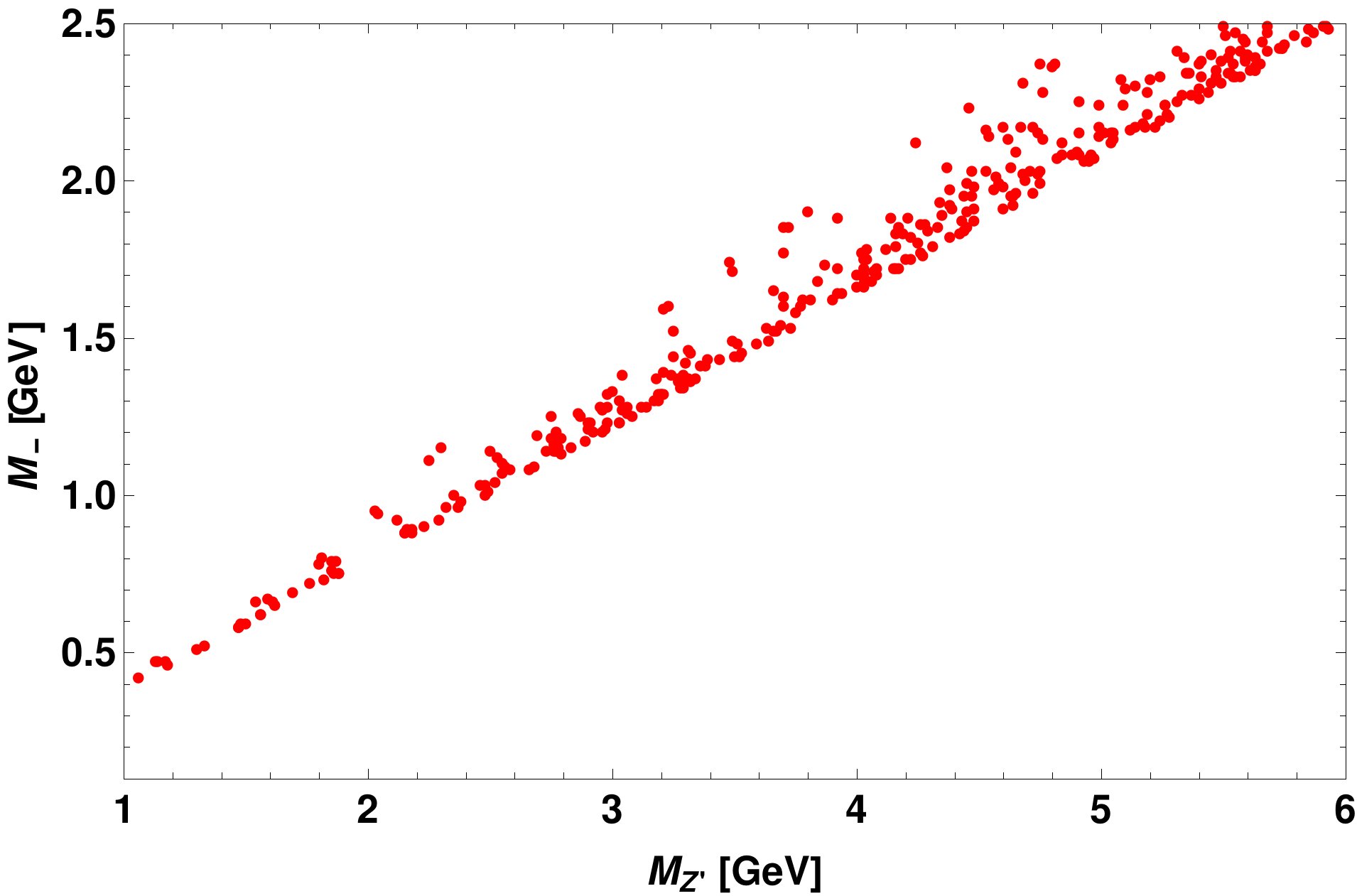}
\caption{Behavior of relic density plotted against DM mass with black horizontal dotted lines denoting the $3\sigma$ range of Planck limit \citep{Aghanim:2018eyx}. Right panel corresponds to the parameter space consistent with $3\sigma$ allowed region of Planck data.}
\label{relic_curve}
\end{center}
\end{figure}
\subsection{Detection prospects}
\begin{figure}[thb]
\begin{center}
\includegraphics[width=0.48\linewidth]{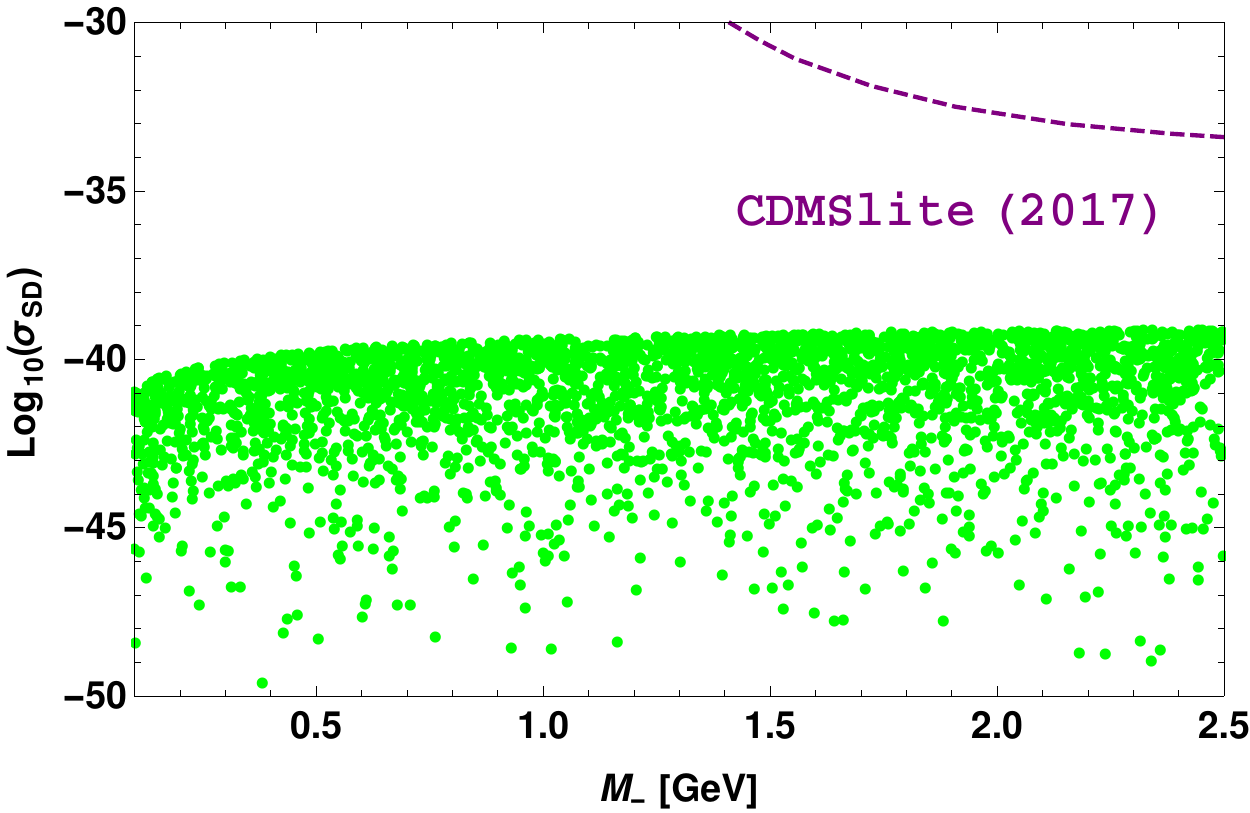}
\caption{Spin-dependent WIMP-proton cross section as a function of DM mass in SLQ portal. Dashed line represent the bound levied by CDMSlite \cite{Agnese:2017jvy}.}
\label{DDscatter}
\end{center}
\end{figure}

Moving to detection paradigm, SLQ portal spin-dependent (SD) cross section can arise from the effective interaction
\begin{equation}
\mathcal{L^{\rm SD}_{\rm eff}} \simeq \frac{y_{qR}^2\cos^2\beta}{4(M_{S_1}^2-M_-^2)} \overline{N_-}\gamma^\mu\gamma^5 N_- \overline{q}\gamma_\mu\gamma^5 q\,.
\end{equation}
The computed cross section is given by \cite{Agrawal:2010fh}
\begin{equation}
\sigma_{\rm SD} = \frac{ \mu_r^2}{\pi} \frac{\cos^4\beta}{(M_{S_1}^2 - M_-^2)^2}\left[y_{dR}^2\Delta_d + y_{sR}^2\Delta_s\right]^2 J_n(J_n+1),
\end{equation}
where $J_n = \frac{1}{2}$ stands for angular momentum, $\mu_r$ represents the reduced mass and the values of quark spin functions $\Delta_{q}$ are provided in \citep{Agrawal:2010fh}. Fig. \ref{DDscatter} projects the SD cross section and it is clear that it is  well below the experimental upper limit from CDMSlite \cite{Agnese:2017jvy}. Furthermore, the WIMP-nucleon cross section in gauge-portal (via $Z$, $Z^\prime$) and scalar-portal (via $H_1$, $H_2$) is found to be very small and insensitive to direct detection experiments.
%
%
\section{Brief comments on neutrino mass}
Neutrino mass can be obtained at one-loop level from the Yukawa interaction with $\eta$ in Eqn. (\ref{Lag}). Assuming $ (M^2_{\eta_r} + M^2_{\eta_i})/{2} = m_0^2$ is much greater than $M^2_{\eta_r} -M^2_{\eta_i} = \lambda''_{H\eta}v^2$, the expression for the radiatively generated neutrino mass \cite{Ma:2006km,Vicente:2015zba} is given by
\begin{equation}
({\cal M}_\nu)_{\beta\gamma} = {\lambda''_{H\eta} v^2  \over32 \pi^{2}} 
\sum_{i=1}^3 {Y_{ \beta i} Y_{\gamma i} \over M_{ii}} \left[ 
{M_{ii}^2 \over m_0^{2} - M_{ii}^{ 2}} +{M_{ii}^4 \over (m_0^{2}-M^2_{ii})^2} \ln { M^2_{ii}\over m_0^{2}  }  \right],\label{nu-mass}
\end{equation}
where $M_{ij} = {\rm diag}(M_{ee},M_-,M_+)$. For $m_0 \sim 1$ TeV, $\lambda_5 \sim 10^{-1}$ and $Y\sim 10^{-4}$, one can obtain neutrino mass at sub-eV scale. It should be noted that as the $L_\mu-L_\tau$ charges of SM leptons and the heavy fermions are same for each generation while the inert doublet is charged zero, the Yukawa matrix $Y$ is diagonal. Hence, the neutrino mass matrix (\ref{nu-mass}) is diagonal by model construction, i.e., ${\cal M}_\nu={\rm diag}(m_1, m_2, m_3)$, thus it can automatically accommodate the observed neutrino mixing angles, i.e., any unitary matrix can be used as the neutrino mixing matrix.

\section{Constraints on new parameters from the flavor sector}

Now, we look forward to constrain the model parameters of LQ and $Z^\prime$ couplings using the branching ratios of  $B \to K \mu \mu$, $B \to K^* \mu \mu$, $B \to X_s \gamma$ decay modes and the recent measurements on lepton non-universality  parameters, $R_{K^{(*)}}$.  

\subsection{$b \to sll$}
The general effective Hamiltonian mediating the $ b \to s l^+ l^-$  transition is given by \cite{Bobeth:1999mk, Bobeth:2001jm} 
\bea
{\cal H}_{\rm eff} &=& - \frac{ 4 G_F}{\sqrt 2} V_{tb} V_{ts}^* \Bigg[\sum_{i=1}^6 C_i(\mu) O_i +\sum_{i=7,9,10} \Big ( C_i(\mu) O_i
+ C_i^\prime(\mu) O_i^\prime \Big )
\Bigg]\;,\label{ham}
\eea
where $G_F$ is the Fermi constant,  $V_{qq^\prime}$ denote the CKM matrix elements, $C_i$'s stand for  the Wilson coefficients evaluated at the renormalized scale $\mu = m_b$ \cite{Hou:2014dza} and the values at NLL ($C_{9,10}$ values are calculated in the NNLL order) are listed in Table \ref{SM-WC}\,.

\begin{table}[htb]
\centering
\begin{tabular}{c c c c c c c c c c}
\hline
\hline
$C_1$~&~$C_2$~&~$C_3$~&~$C_4$~&~$C_5$~&~$C_6$~&~$C_7^{\rm eff}$~&~$C_8^{\rm eff}$~&~$C_9$~&~$C_{10}$ \\
\hline
$-3.001$~&~$1.008$~&~$-0.0047$~&~$-0.0827$~&~$0.0003$~&~$0.0009$~&~$-0.2969$~&~$-0.1642$~&~$4.2607$~&~$-4.2453$~\\
\hline
\hline
\end{tabular}
\caption{The SM Wilson coefficients computed at the scale $\mu=4.6$ GeV \cite{Hou:2014dza}.} \label{SM-WC}
\end{table} 

 Here $O_i$'s represent dimension-six operators responsible for leptonic/semileptonic processes, given as
\bea
O_7^{(\prime)} &=&\frac{e}{16 \pi^2} \Big[\bar s \sigma_{\mu \nu}
\left (m_s P_{L(R)} + m_b P_{R(L)} \right ) b\Big] F^{\mu \nu}, \nn\\
O_9^{(\prime)}&=& \frac{\alpha_{\rm em}}{4 \pi} (\bar s \gamma^\mu P_{L(R)} b)(\bar l \gamma_\mu l)\;,~~~~~~~ O_{10}^{(\prime)}= \frac{\alpha_{\rm em}}{4 \pi} (\bar s \gamma^\mu 
P_{L(R)} b)(\bar l \gamma_\mu \gamma_5 l)\;,
\eea
with $\alpha_{\rm em}$ is  the fine-structure constant, $P_{L,R} = (1\mp \gamma_5)/2$ are the chiral operators and the values of primed Wilson coefficient are zero in the SM, but can be nonzero in the proposed $L_\mu-L_\tau$  model.

\begin{figure}[thb]
\begin{center}
\includegraphics[width=0.24\linewidth]{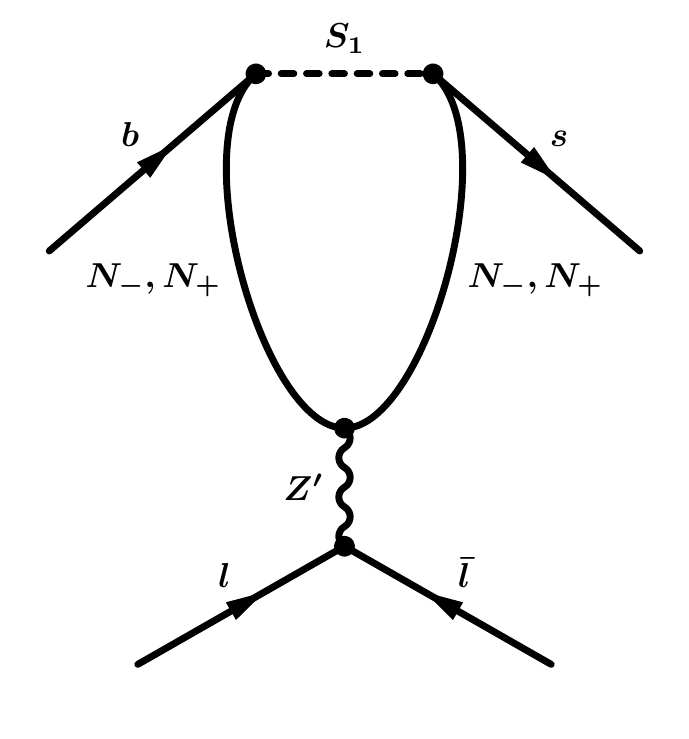}
\includegraphics[width=0.24\linewidth]{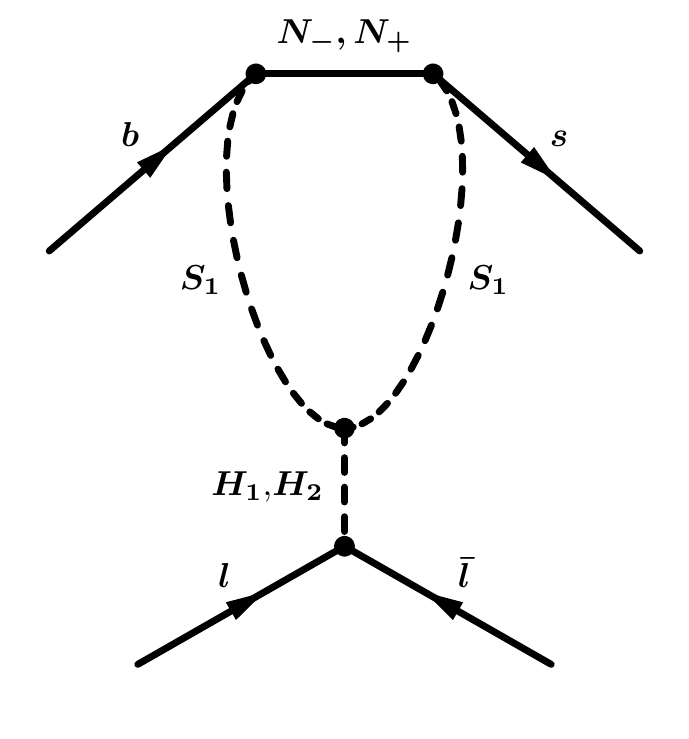}
\includegraphics[width=0.24\linewidth]{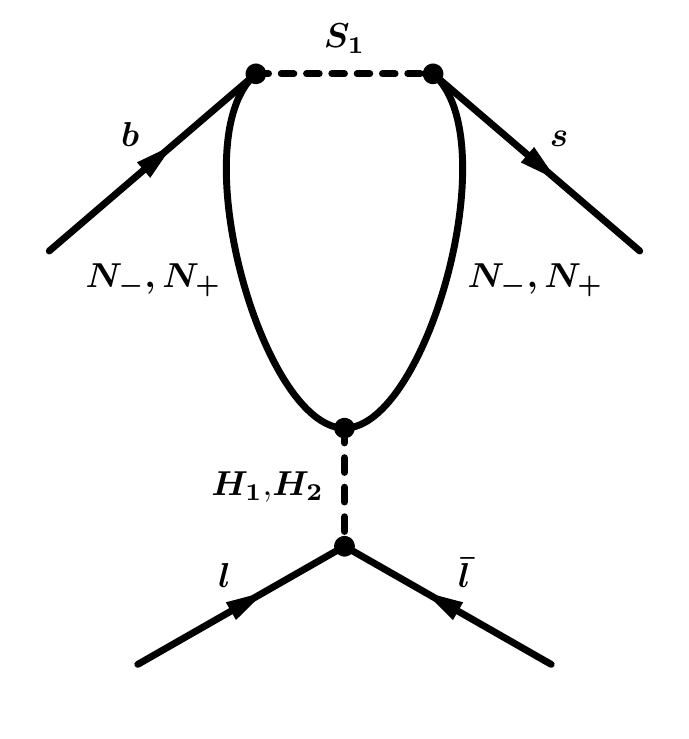}
\caption{One loop penguin diagrams that provide non-zero contribution to  $b \to s ll$ transitions  in the present  model.}
\label{Fig:peng_ll}
\end{center}
\end{figure}

The one loop diagrams that provide non-zero contribution to the rare $b \to sll$ processes can take place via the exchange of $Z^\prime, H_{1,2}$  one-loop penguin diagrams with SLQ and $N_\pm$ particles inside the loop as  shown in Fig.  \ref{Fig:peng_ll}\,.  The loop functions of second and third diagrams have $m_qM_{\pm}/M_{S_1}^2$ factor suppression, hence provide minimal contribution to $b \to sll$ processes. Thus, only the first diagram, mediated via $Z^\prime$ boson will contribute significantly to the $b \to s ll$ channels.

In the presence of $Z^\prime$ exchanging one loop diagram, the transition amplitude of semileptonic $b \to sll$ decay process  is given by \cite{Singirala:2018mio}
\bea \label{amp}
\mathcal{M}=\frac{1}{2^{5}\pi^2}\frac{y_{q R}^2 g_{\mu \tau}^2}{(q^2-M_{Z^\prime}^2)} \mathcal{V}_{sb}(\chi_-, \chi_+) [\bar{u}(p_B)\gamma^\mu (1+\gamma_5)u(p_K))][\bar{v}(p_2)\gamma_\mu u(p_1))],\label{loop}
\eea
which in comparison  with the generalized effective Hamiltonian provides additional  primed Wilson coefficient \cite{Singirala:2018mio}
\bea \label{new-C9p}
C_9^{\prime \rm NP } =\frac{\sqrt{2} }{2^4\pi G_F \alpha_{\rm em} V_{tb}V_{ts}^*}\frac{y_{q R}^2 g_{\mu \tau}^2}{(q^2-{M_{Z^\prime}}^2)}\mathcal{V}_{sb}\left(\chi_-, \chi_+\right)\,.
\eea
to $b \to sll$ process. Here $p_B, p_K$ and $p_{1,2}$ are the four momenta of initial $B$ meson, final $K$ meson and  the charged leptons respectively.  The detailed expression for the loop function $\mathcal{V}_{sb}(\chi_-, \chi_+)$ with $\chi_\pm = M_\pm^2/M_{S_1}^2$ is given in Appendix A  \cite{Baek:2017sew, Hisano:1995cp}. Since there is only $C_9^{\prime \rm NP}$ in this model, the $B_s \to \mu \mu (\tau \tau)$ won't play any role in constraining the new parameters. 

\subsection*{$\boldsymbol{B \to K l^+ l^- }$}

Including the new physics contribution, the  differential branching ratio of $B \to K ll$  process  in terms of $q^2$ is given by \cite{Bobeth:2007dw}
 \bea
 \frac{d {\rm Br}}{d q^2 }= \tau_B \frac{G_F^2 \alpha_{\rm em}^2 |V_{tb}V_{ts}^*|^2}{2^8 \pi^5 M_B^3}\sqrt{\lambda(M_B^2, M_K^2, q^2)} \beta_l f_+^2 \Big ( a_l(q^2)+\frac{c_l(q^2)}{3} \Big )\;,
 \eea
 where
 \bea
 a_l(q^2)&=&  q^2|F_P|^2+ \frac{\lambda(M_B^2, M_K^2, q^2)}{4}
 (|F_A|^2+|F_V|^2) \nn \\ &&+ 2 m_l (M_B^2-M_K^2+q^2) {\rm Re}(F_P F_A^*) +4 m_l^2 M_B^2 |F_A|^2 \;,\nn\\
  c_l(q^2)&=& -   \frac{\lambda(M_B^2, M_K^2, q^2)}{4} \beta_l^2 
 \left (|F_A|^2+|F_V|^2\right ),
 \eea
 with
\bea
F_V & =&\frac{2 m_b}{M_B} C_7^{\rm eff}+ C_9^{\rm eff} +  C_9^{\prime \rm NP}, ~~~~~F_A = C_{10}^{\rm SM},\nn\\
F_P&=&  m_l C_{10}^{\rm SM} \Big[ \frac{M_B^2 -M_K^2}{q^2}\Big(\frac{f_0(q^2)}{f_+(q^2)}-1\Big) -1 \Big]\;,
\eea 
and  
\bea
\lambda(a,b,c) =a^2+b^2+c^2-2(ab+bc+ca),~~~~~\beta_l = \sqrt{1-4 m_l^2/q^2}\;.
\eea
 The detailed expression for $C_{7,9}^{\rm eff}$ Wilson coefficients \cite{Bobeth:2011gi, Grinstein:2004vb, Bobeth:2010wg} are presented in Appendix C.
By using the  lifetime  of $B$ meson and  masses of all the particles from \cite{Zyla:2020zbs}, the $B\to K$ form factors from \cite{Colangelo:1996ay}, the predicted branching ratio values of $B \to K \mu \mu (\tau \tau)$ processes  are 
\bea
&&{\rm Br}(  B^0 \to  K^0 \mu^+ \mu^-)\big |^{\rm SM} =(1.48\pm 0.12)\times 10^{-7}\,,\\
&&{\rm Br}(  B^+ \to  K^+ \mu^+ \mu^-)\big |^{\rm SM} =(1.6\pm 0.13)\times 10^{-7}\,,\\
&&{\rm Br}(B^+ \to K^+ \tau^+ \tau^-)\big |^{\rm SM} =(1.52\pm 0.121)\times 10^{-7}\,,
\eea 
which in comparison with the corresponding experimental data \cite{Zyla:2020zbs}
\bea
&&{\rm Br}( B^0 \to  K^0 \mu^+ \mu^-)\big |^{\rm Expt} = (3.39\pm 0.34)\times 10^{-7}\,,\\
&&{\rm Br}( B^+ \to  K^+ \mu^+ \mu^-)\big |^{\rm Expt} = (4.41\pm 0.22)\times 10^{-7}\,,\\
&&{\rm Br}(B^+ \to K^+ \tau^+ \tau^-)\big |^{\rm Expt} < 2.25\times 10^{-3}\,,
\eea
can put constraints on $y_{qR}, ~g_{\mu \tau},~M_-$ and $M_{Z^\prime}$  parameters. As $Z^\prime$ does not couple to electrons, here we have considered only the channels with  $\mu$ and $\tau$ in the final state. 

\subsection*{$\boldsymbol{B \to K^*  l^+ l^-}$ }

The dilepton invariant mass spectrum for ${B} \rightarrow  K^* l^+ l^-$ decay after integration over all angles; $\theta_l$ (angle between $l^-$ and 
$B$ in the dilepton frame), $\theta_{K^*}$ (angle between $K^-$ and $B$ in the $K^-\pi^+~$ frame) and $\phi$  (angle between the normal of the $K^-\pi^+$ and the dilepton  planes) \cite{Bobeth:2008ij} is given by
 \bea
 \frac{d\Gamma}{dq^2} = \frac{3}{4} \left(J_1 - \frac{J_2}{3}\right), ~~~~~~J_{1,2} = 2J_{1,2}^s + J_{1,2}^c\,,
 \eea
where the detailed  expressions for $J_{1,2}^{s(c)}$ as a function of transversity amplitudes   are given as \cite{Altmannshofer:2008dz},
\bea
 J^s_1 &=& \frac{\left(2+\beta ^2_l\right)}{4}\Bigg[|A_\perp ^L|^2 + |A_\parallel ^L|^2 + \left(L\rightarrow R\right)\Bigg] + \frac{4m^2_l}{q^2} 
{\rm Re}\left(A_\perp ^L A_\perp ^{R^*} + A_\parallel ^L A_\parallel ^{R^*}\right),~~ \nn\\
  J^c_1 &=& |A_0^L|^2 + |A_0^R|^2 +\frac{4m^2_l}{q^2}\Bigg[|A_t|^2 + 2{\rm Re}\left(A_0^L A_0^{R^*}\right)\Bigg], \nn\\
  J^s_2 &=& \frac{\beta^2_l}{4}\left[|A_\perp ^L|^2 + |A_\parallel ^L|^2 + \left(L\rightarrow R\right)\right], \nn\\
  J^c_2 &=& -\beta^2_l\left[|A_0^L|^2 +\left(L\rightarrow R\right)\right],
 \eea
 with
 \bea
 A_i A_j^* = A_{i}^{ L}\left(q^2\right) A^{* L}_{j}\left(q^2\right) + A_{i}^{ R}\left(q^2\right) A^{* R}_{j}\left(q^2\right) 
\hspace{1cm} \left(i,j = 0, \parallel, \perp\right),
 \eea
 in shorthand notation. The  transversity  amplitudes  in terms of the form factors  and (new) Wilson coefficients are given as \cite{Altmannshofer:2008dz}
\bea
A_{\perp L,R}&=& N\sqrt{2  \lambda(M_{K^*}^2, M_B^2, q^2)}\Big [ \left ( (C_9^{\rm eff}+C_9^{\prime \rm NP}) \mp C_{10}^{\rm SM} \right )
\frac{V(q^2)}{M_B+M_{K^*}} + \frac{2 m_b}{q^2} C_7 T_1(q^2) \Big]\;,\nn\\
A_{\para L,R}&=& -N \sqrt{2} (M_B^2 -M_{K^*}^2)\Big[ \left ((C_9^{\rm eff}-C_9^{\prime \rm NP}) \mp C_{10}^{\rm SM} \right ) \frac{A_1(q^2)}{M_B - M_{K^*}} + \frac{2 m_b}{q^2}C_7 T_2(q^2) \Big]\;,\nn\\
A_{0 L,R}& =& - \frac{N}{2 M_{K^*} \sqrt{s}} \Big[ \left (C_9^{\rm eff}-C_9^{\prime \rm NP}) \mp C_{10}^{\rm SM} \right ) \nn\\
&& \times \left ( (M_B^2 -M_{K^*}^2 -q^2)(M_B + M_{K^*}) A_1(q^2)-\lambda(M_{K^*}^2, M_B^2, q^2) \frac{A_2(q^2)}{M_B + M_{K^*}} \right )\nn\\
&&+ 2 M_B C_7 \left ( (M_B^2 +3 M_{K^*}^2 -q^2) T_2(q^2) - \frac{\lambda(M_{K^*}^2, M_B^2, q^2)}{M_B^2 - M_{K^*}^2} \right ) \Big]\;,\nn\\
A_t&=& 2N \sqrt{\frac{\lambda(M_{K^*}^2, M_B^2, q^2)}{q^2}}   C_{10}^{\rm SM}  A_0(q^2),
\eea
where
\bea
N= V_{tb}V_{ts}^* \left [ \frac{G_F^2 \alpha_{\rm em}^2}{3 \cdot 2^{10} \pi^5 M_B^3} q^2 \beta_l \sqrt{\lambda(M_{K^*}^2, M_B^2, q^2)} \right ]^{1/2}\;. 
\eea
With the use of the mass and lifetime of particles from Particle Data Group \cite{Zyla:2020zbs} and the form factor from \cite{Ball:2004rg}, the predicted branching ratios of $B \to K^* \mu \mu (\tau \tau)$ are 
\bea
&&{\rm Br}(  B^0 \to  K^{* 0} \mu^+ \mu^-)\big |^{\rm SM} =(1.967\pm 0.158)\times 10^{-8}\,,\\
&&{\rm Br}(B^+ \to K^{* +} \mu^+ \mu^-)\big |^{\rm SM} =(1.758\pm 0.141)\times 10^{-8}\,,
\eea 
which in correlation  with the corresponding experimental data \cite{Zyla:2020zbs}
\bea
&&{\rm Br}( B^0 \to  K^{* 0} \mu^+ \mu^-)\big |^{\rm Expt} = (9.4\pm 0.5)\times 10^{-7}\,,\\
&&{\rm Br}(B^+ \to K^{*+} \mu^+ \mu^-)\big |^{\rm Expt} < 5.9\times 10^{-7}\,,
\eea
can  constrain the  $y_{qR}, ~g_{\mu \tau},~M_-$ and $M_{Z^\prime}$  parameters.
\subsection*{$\boldsymbol{R_{K^{(*)}}}$}

Using the full Run-I and Run-II data set, recently the LHCb Collaboration has updated the lepton non-universality $R_K$ parameter in the $q^2\in [1,6]$~${\rm GeV}^2$ \cite{Aaij:2021vac}
\bea \label{Eqn:RK-Exp-new}
R_K^{\rm LHCb21} \ = \ \frac{{\rm Br}(B^+ \to K^+ \mu^+ \mu^-)}{{\rm Br}(B^+ \to K^+ e^+ e^-)}=\ 0.846^{+0.042+0.013}_{-0.039-0.012}\
\eea
which is pretty much precise than the previous result \cite{Aaij:2019wad}
\bea \label{Eqn:RK-Exp-old}
R_K^{\rm LHCb19} \ = \ 0.846^{+0.060+0.016}_{-0.054-0.014}\,,
\eea
(where the first uncertainty is statistical and the second one is systematic),
giving rise to the disagreement of $3.1\sigma$ from the SM prediction ~\cite{Bobeth:2007dw, Bordone:2016gaq} 
\bea \label{Eqn:RK-SM}
R_K^{\rm SM} \ = \ 1.0003\pm 0.0001\,.
\eea

Equivalently, the recent measurements by the LHCb experiment on $R_{K^{*}}$ ratio in two bins of low-$q^2$ regions~\cite{Aaij:2017vbb}:
\bea
R_{K^*}^{\rm LHCb}& \ = \ & \begin{cases}0.660^{+0.110}_{-0.070}\pm 0.03 \qquad q^2\in [0.045, 1.1]~{\rm GeV}^2 \, , \\ 
0.69^{+0.11}_{-0.07}\pm 0.05 \qquad q^2\in [1.1,6.0]~{\rm GeV}^2 \, .
\end{cases}
\eea 
 have respectively $2.1\sigma$ and $2.5\sigma$ deviations from their corresponding SM values~\cite{Capdevila:2017bsm}:
\bea
R_{K^*}^{\rm SM} \ = \  \begin{cases} 0.92\pm 0.02 \qquad q^2\in [0.045, 1.1]~{\rm GeV}^2 \, , \\  
1.00\pm 0.01\qquad q^2\in [1.1,6.0]~{\rm GeV}^2 \, .
\end{cases}
\eea
Additionally, though the Belle Collaboration has also announced measurements on $R_K$~\cite{Abdesselam:2019lab} and 
$R_{K^*}$~\cite{Abdesselam:2019wac} in several other bins, their results  have comparatively larger uncertainties. The $R_{K^{(*)}}$ ratios can constrain all the four new parameters.

\subsection{$b \to s \gamma$}
\begin{figure}[thb]
\begin{center}
\includegraphics[width=0.4\linewidth]{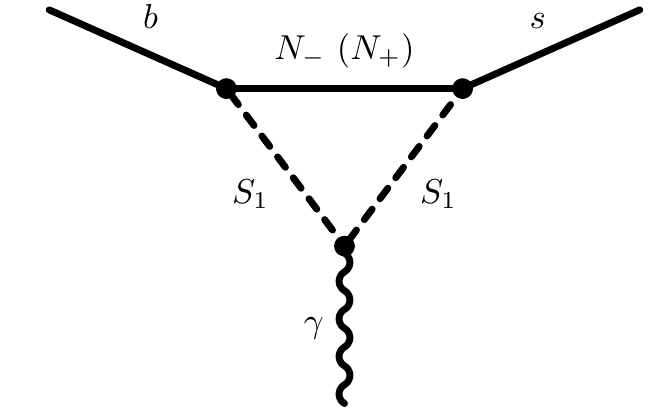}
\caption{ One loop diagram of $ b \to s \gamma$  processes in the proposed model.}
\label{Fig:bsgam}
\end{center}
\end{figure}

The $b \to s \gamma$ transition can occur at one loop level as shown in Fig. \ref{Fig:bsgam}, where leptoquark and $N_\pm$ were driving in the loop. 

\subsection*{$ \boldsymbol{B \to X_s \gamma}$}

In the presence of NP, the branching ratio of $ B \to X_s \gamma$ decay process induced by the $b \to s \gamma$ transition is
\bea \label{tot-bsgamma}
{\rm Br}( B \to X_s \gamma) ={\rm Br}( B \to X_s \gamma)\big |^{\rm SM} \Bigg ( 1+\frac{C_7^{\gamma \prime \rm NP}}{C_7^{\gamma \rm SM}} \Bigg )^2,
\eea
where
\bea \label{c7new}
C_7^{\gamma \prime \rm NP}= -\frac{\sqrt{2}/3}{8 G_F V_{tb} V_{ts}^*}\frac{y_{qR}^2}{M_{S_1}^2} \Big( J_1(\chi_-)\cos^2 \alpha +J_1 (\chi_+)\sin^2 \alpha \Big),
\eea
with the loop functions $J_1(\chi_{\pm})$   \cite{Baek:2017sew}
\bea
J_1(\chi_\pm)=\frac{1-6\chi_\pm+3\chi_\pm^2+2\chi_\pm^3-6\chi_\pm^2 \log \chi_\pm}{12 (1-\chi_\pm)^4}\,.
\eea  
The predicted SM branching ratio of $ B \to X_s \gamma$ process  \cite{Misiak:2015xwa}
\bea \label{SM-bsgamma}
{\rm Br}( B \to X_s \gamma)\big |^{\rm SM}_{E_\gamma > 1.6~{\rm GeV}}=(3.36\pm 0.23)\times 10^{-4}\,,
\eea
in relation with the corresponding experimental limit \cite{Amhis:2016xyh}
\bea \label{Exp-bsgamma}
{\rm Br}( B \to X_s \gamma)\big |^{\rm Expt}_{ E_\gamma > 1.6~{\rm GeV}}=(3.32\pm 0.16)\times 10^{-4}\,,
\eea
will impose constrain on the $y_{qR},~M_-$ parameters. 

\begin{figure}[htb]
\centering
\includegraphics[width=0.48\linewidth]{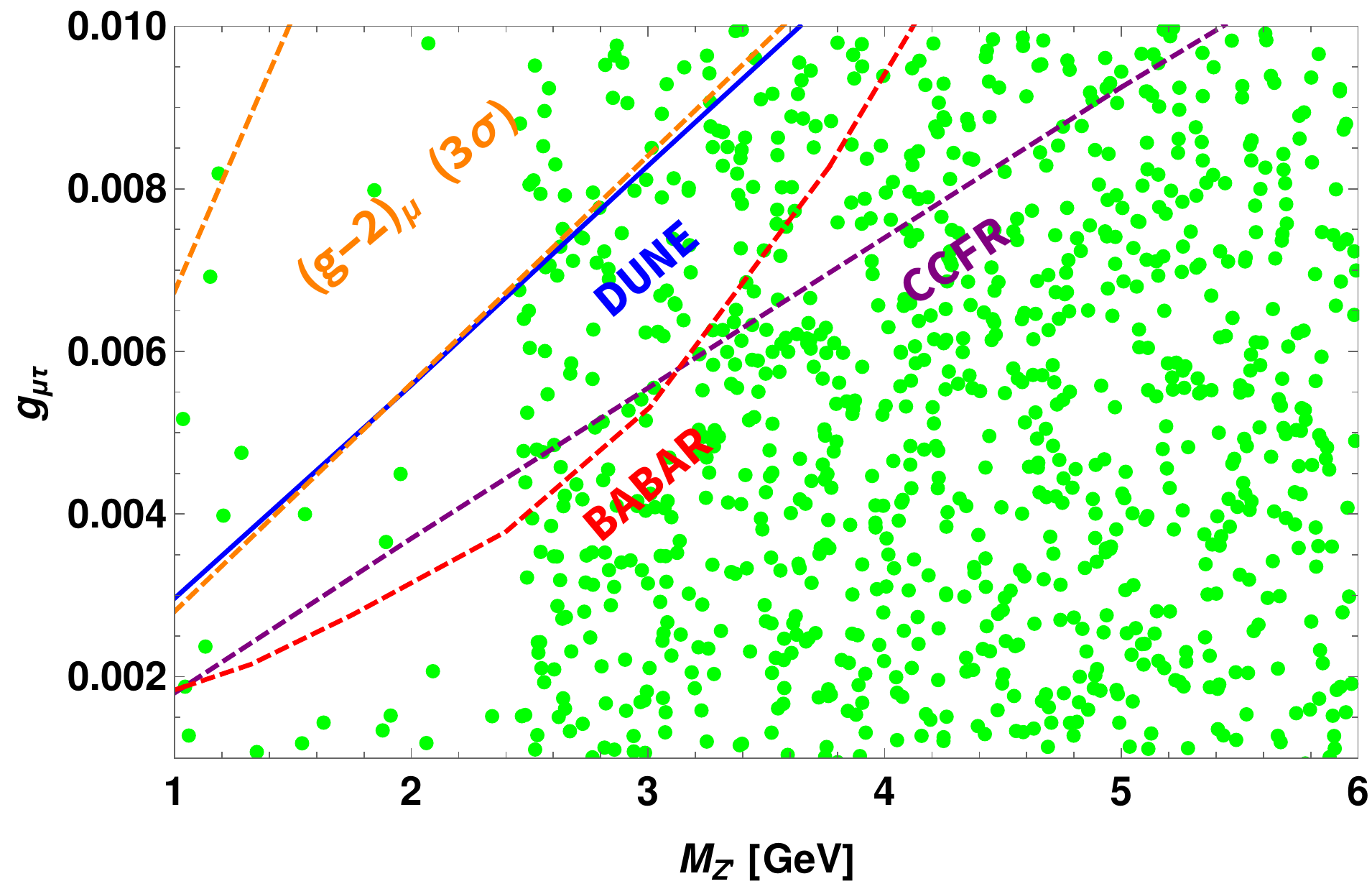}
\quad
\includegraphics[width=0.47\linewidth]{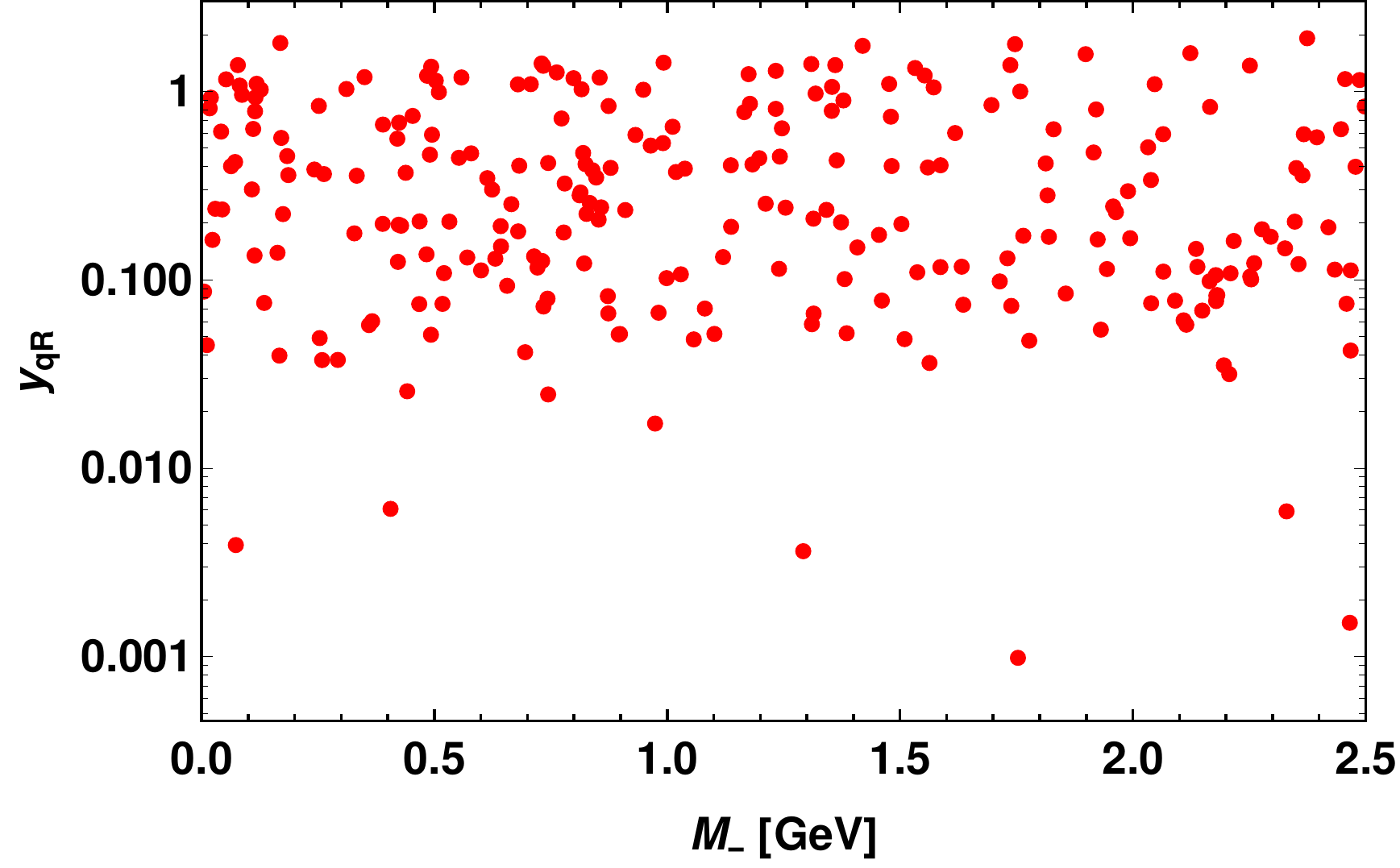}
\caption{Constraints on $M_Z^\prime-g_{\mu \tau}$ (left panel) and $M_--y_{qR}$ (right panel), obtained by using the branching ratios of $b \to s ll$, $b \to s \gamma$ processes and the $R_{K^{(*)}}$ parameters. Lines in left panel denote the $3\sigma$ consistent region of muon $g-2$ \cite{Abi:2021gix}, exclusion limit from BABAR \cite{TheBABAR:2016rlg}, neutrino trident bounds from CCFR \cite{Mishra:1991bv,Altmannshofer:2014pba} and DUNE \cite{Altmannshofer:2019zhy}.} \label{Fig:Constraint}
\end{figure}

Due to the absence of $Z^\prime \mu \tau$ coupling in this model, the lepton flavor violating processes like $B \to K^{(*)} \mu \tau$, $\tau \to \mu \gamma$ and $\tau \to 3\mu$ are not allowed, thus couldn't put bound on new parameters. Using Br($B \to K^{(*)}ll$) and $R_{K^{(*)}}$ observables, we show the $g_{\mu\tau},~M_{Z^\prime}$ allowed parameters space in the left panel of Fig. \ref{Fig:Constraint}\,. The region consistent with muon anomalous magnetic moment data is fully factored out by the constraint from the neutrino trident production. The validity of the effective field theory description in the present model lies above electroweak scale i.e., in the range $\sim (300-3000)$ GeV. It is indeed the favorable region of the ratio $M_{Z^\prime}/2g_{\mu\tau}$ and the lower values of this ratio is found to be excluded by various experimental searches, as projected in the plot.
The $y_{qR},~M_-$ allowed region consistent with the Br($B \to K^{(*)}ll$), Br($B \to X_s \gamma$)  and $R_{K^{(*)}}$ observables are presented in the right panel of Fig. \ref{Fig:Constraint}\,.  The allowed range of all the four new parameters consistent with flavor phenomenology is given in Table \ref{Tab:Constraint}\,.
 \begin{table}[htb]
 \centering
 \begin{tabular}{|c|c|c|c|c|c|c|}
 \hline
~Parameters~& ~$y_{qR}$~&~$g_{\mu \tau}$~&~$M_-$ (GeV)~&~$M_{Z^\prime}$ (GeV)\\
 \hline
 ~Allowed range~&~$0-2.0$~&~$0-0.01$~&~$0-2.5$~&$1-6$\\
 \hline
 \end{tabular}
 \caption{The allowed regions of $y_{qR}, ~g_{\mu \tau},~M_-$ and $M_{Z^\prime}$ parameters.}\label{Tab:Constraint}
 \end{table}

\section{Footprints on $b \to s +\displaystyle {\not} E$ decay modes}

The SM treats neutrinos as the only carrier of missing energy in $b \to s$ transitions, i.e., the $b \to s +$ missing energy can be described by the $b \to s \nu \bar \nu$ decay modes in the SM. Here we consider the lepton flavor conserving $b \to s \nu_l \bar \nu_l$ processes. In the SM, the  general effective Hamiltonian responsible for the $b \to s \nu_l \bar{\nu_l}$ transition is given by  \cite{Altmannshofer:2009ma} 
\begin{equation}
\mathcal{H}_{eff} = \frac{-4G_F}{\sqrt{2}} V_{tb}V_{ts}^*\left(C^\nu_L \mathcal{O}^\nu_L +C^\nu_R \mathcal{O}^\nu_R \right) + h.c., \label{nu1}
\end{equation}
where 
\begin{equation}
\mathcal{O}^\nu_L = \frac{\alpha_{\rm em}}{4 \pi} \left(\bar{s}_R\gamma_\mu  b_L \right) \left(\bar{\nu} \gamma^\mu \left(1-\gamma_5\right)\nu\right),
 \hspace{1cm} \mathcal{O}^\nu_R = \frac{\alpha_{\rm em}}{4 \pi} \left(\bar{s}_L \gamma_\mu  b_R \right) \left(\bar{\nu} \gamma^\mu \left(1-\gamma_5\right)\nu\right),
\label{nu-ham}
\end{equation}
are the dimension-6 current-current  operators with 
\bea
 C^\nu_L = -X(x_t)/\sin^2\theta_w\;, ~~~~X(x_t) = X_0(x_t)+ \frac{\alpha_s}{4\pi} X_1(x_t),\label{cl}
\eea
being the SM Wilson coefficient. The expressions for ($X_{0,1}(x_t)$ loop functions  can be found in Ref. \cite{Misiak:1999yg, Buchalla:1998ba}). In the SM, the  $C^\nu_R$ Wilson coefficient is absent but can be  generated in some new physics models.

\begin{figure}[thb]
\begin{center}
\includegraphics[width=0.24\linewidth]{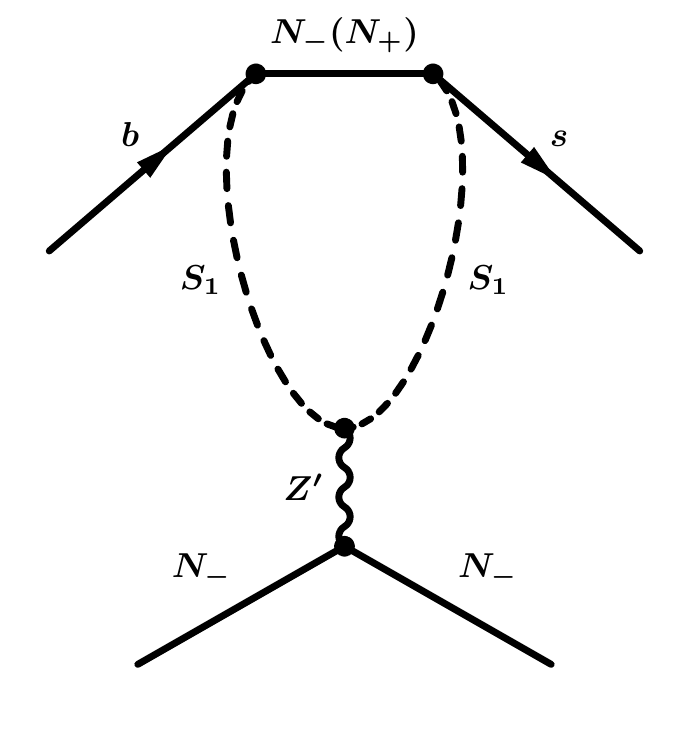}
\includegraphics[width=0.24\linewidth]{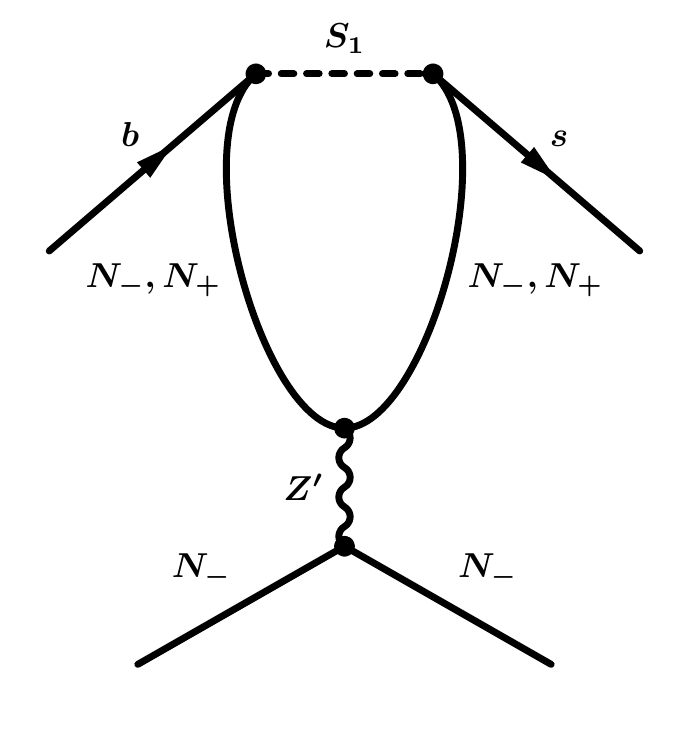}
\includegraphics[width=0.24\linewidth]{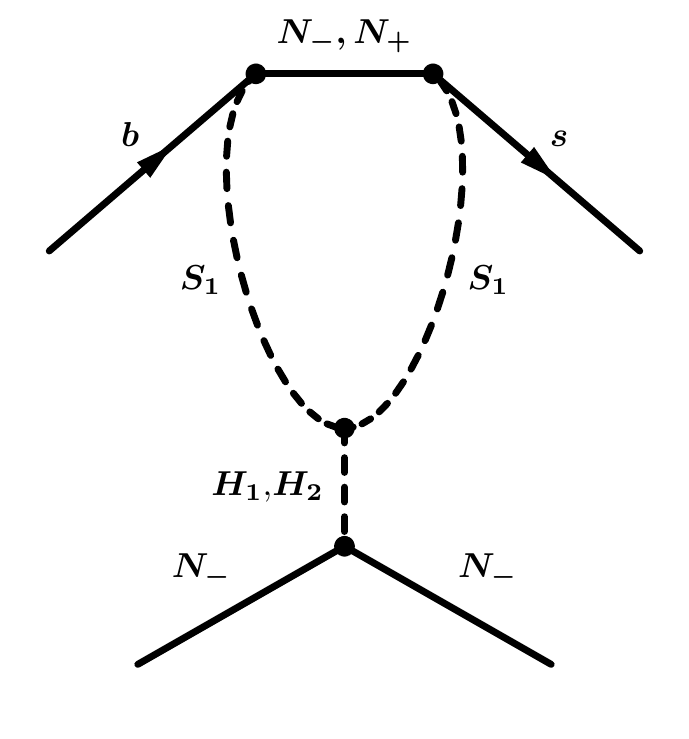}
\includegraphics[width=0.24\linewidth]{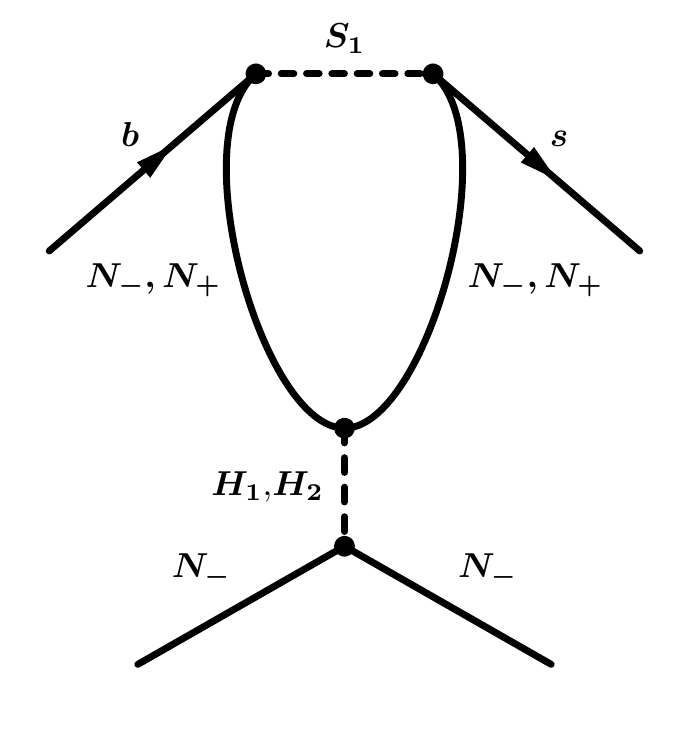}
\includegraphics[width=0.24\linewidth]{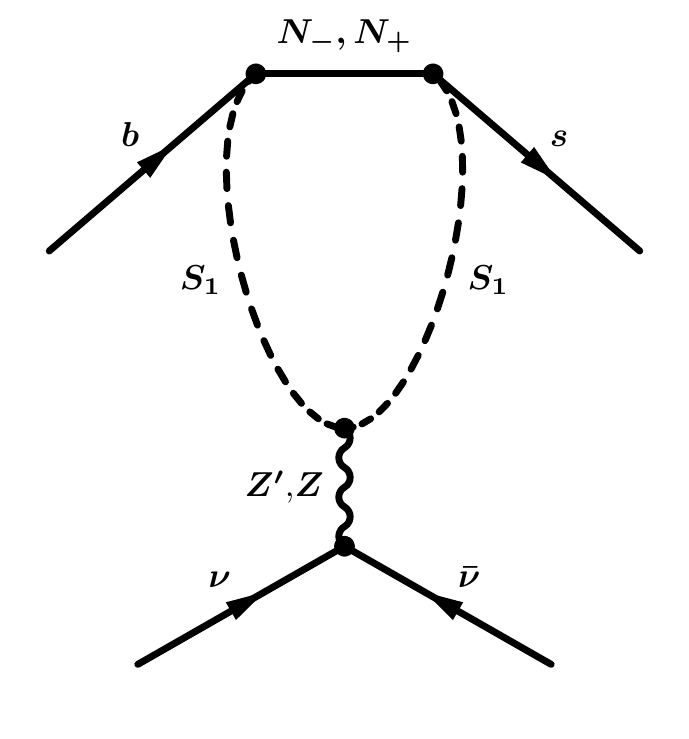}
\includegraphics[width=0.24\linewidth]{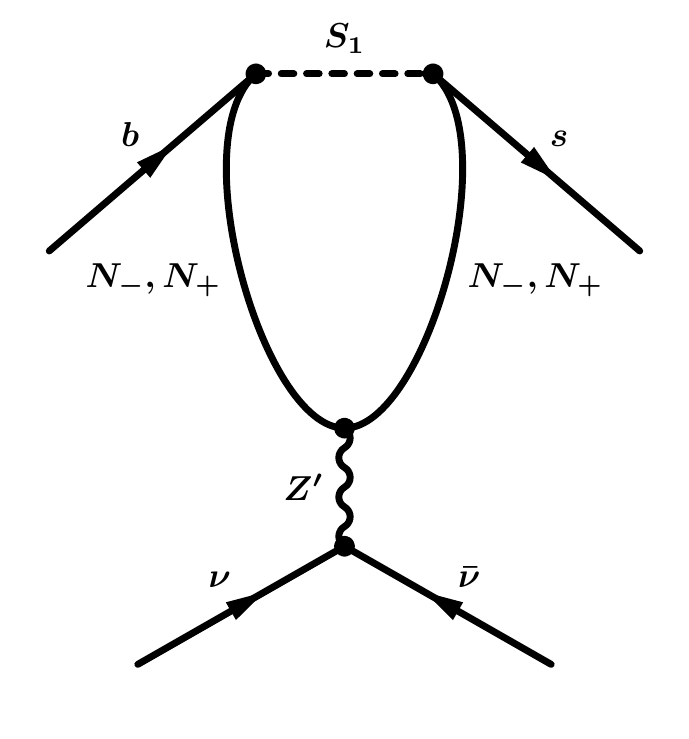}
\caption{One loop penguin diagrams that provide non-zero contribution for $b \to s +$ missing energy decay modes.}
\label{Fig:peng_missing}
\end{center}
\end{figure}

In our model, all possible diagrams contributing to $b \to s +\displaystyle {\not} E$ processes are shown in Fig. \ref{Fig:peng_missing}\,. The first, third, fourth and fifth  diagrams  in  Fig. \ref{Fig:peng_missing}, will be suppressed by the factor $m_q M_{\pm} /M_{S_1}^2$. In sixth diagram, the contributions of muon-neutrino and tau-neutrino cancel with each other in the leading order of NP due to their equal and opposite $L_\mu-L_\tau$ charge. Thus, only the $Z^\prime$ exchanging second diagram  will contribute to $b \to s +\displaystyle {\not} E$ processes in addition to the SM $b \to s \nu_l \bar \nu_l$ modes. 

\subsection{$B \to K+\displaystyle {\not} E$}

The total branching ratio of $B \to K + \displaystyle {\not} E$  decay is the sum of the branching ratio of semileptonic  $B \to K $ decay with SM  neutrino in the final state and  the branching ratio of $B \to K  N_- N_-$ decay, i.e.,
\bea \label{Br-BKs}
{\rm Br} (B \to K + \displaystyle{\not} E) = {\rm Br}(B \to K  \nu \bar \nu) + {\rm Br}(B \to K  N_- N_-).
\eea
The branching ratio of $B \to K \nu_l \bar \nu_l$ decay process in the SM is given by  \cite{Altmannshofer:2009ma}
\bea
\frac{d{\rm Br}}{ds_B} = \tau_B \frac{G_F^2\alpha^2}{256\pi^5} |V_{ts}^* V_{tb}|^2 M_B^5 \lambda^{3/2}(s_B,\tilde{M}_K^2,1) |f_+^K(s_B)|^2 |C_L^\nu |^2,
\label{BKnunu} 
\eea
where  $\tilde{M}_K = M_K/M_B,~ s_B = s/M_B^2$.
The predicted branching ratio values of $B^{+(0)} \to K^{+(0)} \nu_l \bar \nu_l $ by using input values from \cite{Zyla:2020zbs} and the corresponding experimental limits are tabulated in Table \ref{Tab:bsnunu-SM}\,. 

The transition amplitude of $B  \to K N_- N_-$ process from the $Z^\prime$ exchanging one loop penguin diagram is 
\bea \label{bsNN-amp}
\mathcal{M}&=&\frac{1}{2^{5}\pi^2}\frac{y_{b R}^2 g_{\mu \tau}^2}{q^2-M_{Z^\prime}^2} \mathcal{V}_{sb}(\chi_-, \chi_+) [\bar{u}(p_B)\gamma^\mu (1+\gamma_5)u(p_K))][\bar{v}(p_2)\gamma_\mu u(p_1))],\nn\\
&=& C^{\rm NP}(q^2) [\bar{u}(p_B)\gamma^\mu (1+\gamma_5)u(p_K))][\bar{v}(p_2)\gamma_\mu u(p_1))]
\eea
where 
\bea
C^{\rm NP}(q^2)=\frac{1}{2^{5}\pi^2}\frac{y_{q R}^2 g_{\mu \tau}^2 \cos 2\beta \cos \alpha \sec \chi}{q^2-M_{Z^\prime}^2} \mathcal{V}_{sb}(\chi_-, \chi_+)\,,
\eea
and $p_B(p_K)$ is the four momenta of $B(K)$ meson, $p_{1,2}$ are the momenta of $N_-$ fermion. 

Using (\ref{bsNN-amp})\,, the branching ratio of $B \to K N_- N_-$ decay mode is given by 
\bea
 \frac{d {\rm Br}}{d q^2 }= \tau_B \frac{1}{2^7 \pi^3 M_B^3}\sqrt{\lambda(M_B^2, M_K^2, q^2)} \beta_N f_+^2 \Big ( a_l(q^2)+\frac{c_l(q^2)}{3} \Big )\;,~~~~
 \eea
 where
 \bea
 a_l(q^2)&=&  q^2|F_P|^2+ \left( \frac{\lambda(M_B^2, M_K^2, q^2)}{4}+4 M_-^2 M_B^2\right) |F_A|^2+ 2 M_- (M_B^2-M_K^2+q^2) {\rm Re}(F_P F_A^*) \;,\nn\\
  c_l(q^2)&=& -   \frac{\lambda(M_B^2, M_K^2, q^2)}{4} \beta_N^2 |F_A|^2,
 \eea
 with
\bea
F_A = C^{\rm NP}(q^2),~~F_P=  M_- C^{\rm NP}(q^2) \Big[ \frac{M_B^2 -M_K^2}{q^2}\Big(\frac{f_0(q^2)}{f_+(q^2)}-1\Big) -1 \Big], ~~\beta_N=\sqrt{1-4M_-^2/q^2}\,.~~~
\eea

 \subsection*{$\boldsymbol{B \to K^* N_-N_-}$}
 
 Similary, the branching ratio of $B \to K^*+\displaystyle {\not} E$ is given by
 \bea \label{Br-BKstar}
{\rm Br} (B \to K^* + \displaystyle{\not} E) = {\rm Br}(B \to K^*  \nu \bar \nu) + {\rm Br}(B \to K^*  N_- N_-).
\eea
The decay rate of $B \to K^* \nu \bar \nu$ decay mode in terms of $s_B$ and $\cos\theta$  is given as \cite{Altmannshofer:2009ma, Kim:1999waa}
\bea\label{B-Kstar-decay}
\frac{d^2 \Gamma}{ds_B d \cos \theta} = \frac{3}{4} \frac{d \Gamma_T}{ds_B } \sin^2 \theta + \frac{3}{2} \frac{d \Gamma_L}{ds_B } \cos^2 \theta,
\eea
where the longitudinal and transverse polarization decay rates are
\bea
\frac{d \Gamma_L}{ds_B } = 3 M_B^2 |A_0|^2,  \hspace{2cm} 
\frac{d \Gamma_T}{ds_B } = 3 M_B^2 (|A_\perp|^2  + |A_\parallel|^2),
\eea
with the transversity amplitudes, $A_{\perp, \parallel, 0}$ in terms of form factor and Wilson coefficients  are defined as 
\bea
&&A_\perp(s_B) = 2N\sqrt{2} \lambda^{1/2} (1,\tilde{M}_{K*}^2, s_B) C_L^\nu  \frac{V(s_B)}{(1+\tilde{M}_{K^*})}\,,\nn\\
&& A_\parallel (s_B)  = -2N \sqrt{2} (1+\tilde{M}_{K^*}) C_L^\nu A_1 (s_B)\,,\nn\\
&&A_0 (s_B) = -\frac{N C_L^\nu }{\tilde{M}_{K^*}\sqrt{s_B} } \left[ (1-\tilde{M}_{K^*}^2 - s_B) (1 + \tilde{M}_{K^*}) A_1 (s_B)
 - \lambda(1,\tilde{M}_{K^*}^2, s_B) \frac{A_2 (s_B)}{1 + \tilde{M}_{K^*}} \right],~~~
\eea
where $\tilde{M}_{K^*} = M_{K^*}/M_B$ and
\bea
N = V_{tb} V_{ts}^* \left[ \frac{G_F^2 \alpha^2 M_B^3}{3 \cdot 2^{10} \pi^5 } s_B \lambda^{1/2} (1,\tilde{M}_{K^*}^2, s_B) \right]^{1/2}.
\eea
The same expression can be used for $B_s \to \phi \nu_l \bar \nu_l$. Now using all the required input values from \cite{Zyla:2020zbs}, the $B_{(s)} \to K^*(\phi)$ form factor from \cite{Ball:2004rg},  the branching ratio of $B_{(s)} \to K^*(\phi) \nu \bar \nu$ and their corresponding experimental limits are presented in Table \ref{Tab:bsnunu-SM}\,.

\begin{table}[htb]
\begin{center}
\begin{tabular}{| c | c | c |}
\hline
 Decay process & BR in the SM&  Experimental limit \cite{Zyla:2020zbs} \\
\hline
\hline
$B^0 \to K^0 \nu_l  \bar \nu_l$ &   $ (4.53\pm 0.267) \times 10^{-6}$ & $\textless 2.6 \times 10^{-5}$\\
 
 $B^+ \to K^+ \nu_l  \bar \nu_l$ &   $ (4.9\pm 0.288) \times 10^{-6}$ & $\textless 1.6 \times 10^{-5}$\\
 
 $B^0 \to K^{* 0} \nu_l  \bar \nu_l$ & $(9.48\pm 0.752) \times 10^{-6}$ & $\textless 1.8 \times 10^{-5}$\\

$B^+ \to K^{* +} \nu_l  \bar \nu_l$ & $ (1.03\pm 0.06) \times 10^{-5}$ & $\textless 4.0 \times 10^{-5}$\\

$B_s \to \phi \nu_l  \bar \nu_l$ & $ (1.2 \pm 0.07) \times 10^{-5}$ & $\textless 5.4 \times 10^{-3}$\\

\hline
\end{tabular}
\end{center}
\caption{The  predicted branching ratios of $B^{+(0)} \to M^{+(0)}  \nu \bar \nu$  decays in the SM, where  $M  =K^{(*)},   \phi$. Their corresponding experimental limits are also noted down in this table. }\label{Tab:bsnunu-SM}
\end{table}
 
 The decay rate of $B \to K^* N_- N_-$ process is given by
 \bea
 \frac{d\Gamma}{dq^2} = \frac{3}{4} \left(J_1 - \frac{J_2}{3}\right), ~~~~~~J_{1,2} = 2J_{1,2}^s + J_{1,2}^c\,,
 \eea
where 
\bea
 J^s_1 &=& \frac{\left(2+\beta ^2_N\right)}{4}\Bigg[|A_\perp ^L|^2 + |A_\parallel ^L|^2 + \left(L\rightarrow R\right)\Bigg] + \frac{4m^2_N}{q^2} 
{\rm Re}\left(A_\perp ^L A_\perp ^{R^*} + A_\parallel ^L A_\parallel ^{R^*}\right),~~ \nn\\
  J^c_1 &=& |A_0^L|^2 + |A_0^R|^2 +\frac{4M^2_-}{q^2}\Bigg[|A_t|^2 + 2{\rm Re}\left(A_0^L A_0^{R^*}\right)\Bigg], \nn\\
  J^s_2 &=& \frac{\beta^2_l}{4}\left[|A_\perp ^L|^2 + |A_\parallel ^L|^2 + \left(L\rightarrow R\right)\right], \nn\\
  J^c_2 &=& -\beta^2_l\left[|A_0^L|^2 +\left(L\rightarrow R\right)\right], 
 \eea
 with
 the  transversity  amplitudes  in terms of the form factors  and new functions are given as 
\bea
A_{\perp L,R}&=& \mp N_2 C^{\rm NP}(q^2) \sqrt{2  \lambda(M_{K^*}^2, M_B^2, q^2)}
\frac{V(q^2)}{M_B+M_{K^*}} \;,\nn\\
A_{\para L,R}&=& \pm  N_2 C^{\rm NP}(q^2) \sqrt{2} (M_B^2 -M_{K^*}^2)\frac{A_1(q^2)}{M_B - M_{K^*}}\;,\nn\\
A_{0 L,R}& =& \pm \frac{N_2 C^{\rm NP}(q^2) }{2 M_{K^*} \sqrt{s}} 
 \left ( (M_B^2 -M_{K^*}^2 -q^2)(M_B + M_{K^*}) A_1(q^2)-\lambda(M_{K^*}^2, M_B^2, q^2) \frac{A_2(q^2)}{M_B + M_{K^*}} \right )\;,\nn\\
A_t&=& 2N_2 C^{\rm NP}(q^2) \sqrt{\frac{\lambda(M_{K^*}^2, M_B^2, q^2)}{q^2}}  A_0(q^2),
\eea
and 
\bea
N_2= \left [ \frac{1}{3 \cdot 2^{9} \pi^3 M_B^3} q^2 \beta_N \sqrt{\lambda(M_{K^*}^2, M_B^2, q^2)} \right ]^{1/2}\;. 
\eea

For numerical estimation, we took the required input parameters like mass of particles, lifetime of $B_{(s)}$ meson, CKM parameters from \cite{Zyla:2020zbs} and the $B_{(s)}\to K^*(\phi)$ form factors from \cite{Ball:2004rg}\,. We have taken two different benchmark values of all the four new parameters, which are allowed by both the DM and flavor phenomenology, as presented in Table \ref{Tab:Benchmark}\,.
\begin{table}[htb]
\centering
\begin{tabular}{|c|c|c|c|c|c|c|}
\hline
Benchmark~&~ $y_{qR}$~&~$g_{\mu\tau}$~&~$M_-$ (GeV)~&~$M_{Z^\prime}$ (GeV)~\\
\hline \hline
~Benchmark-I~&$2.0$~&~$0.002$~&~$1.7$~&~$4$~\\
~Benchmark-II~&$2.0$~&~$0.008$~&~$1.8$~&~$4.8$~\\
\hline
\end{tabular}
\caption{Benchmark values of $y_{qR},~M_-,~g_{\mu\tau}$ and $M_{Z^\prime}$ parameters used in our analysis. }
\label{Tab:Benchmark}
\end{table}

\begin{figure}[thb]
\begin{center}
\includegraphics[width=0.48\linewidth]{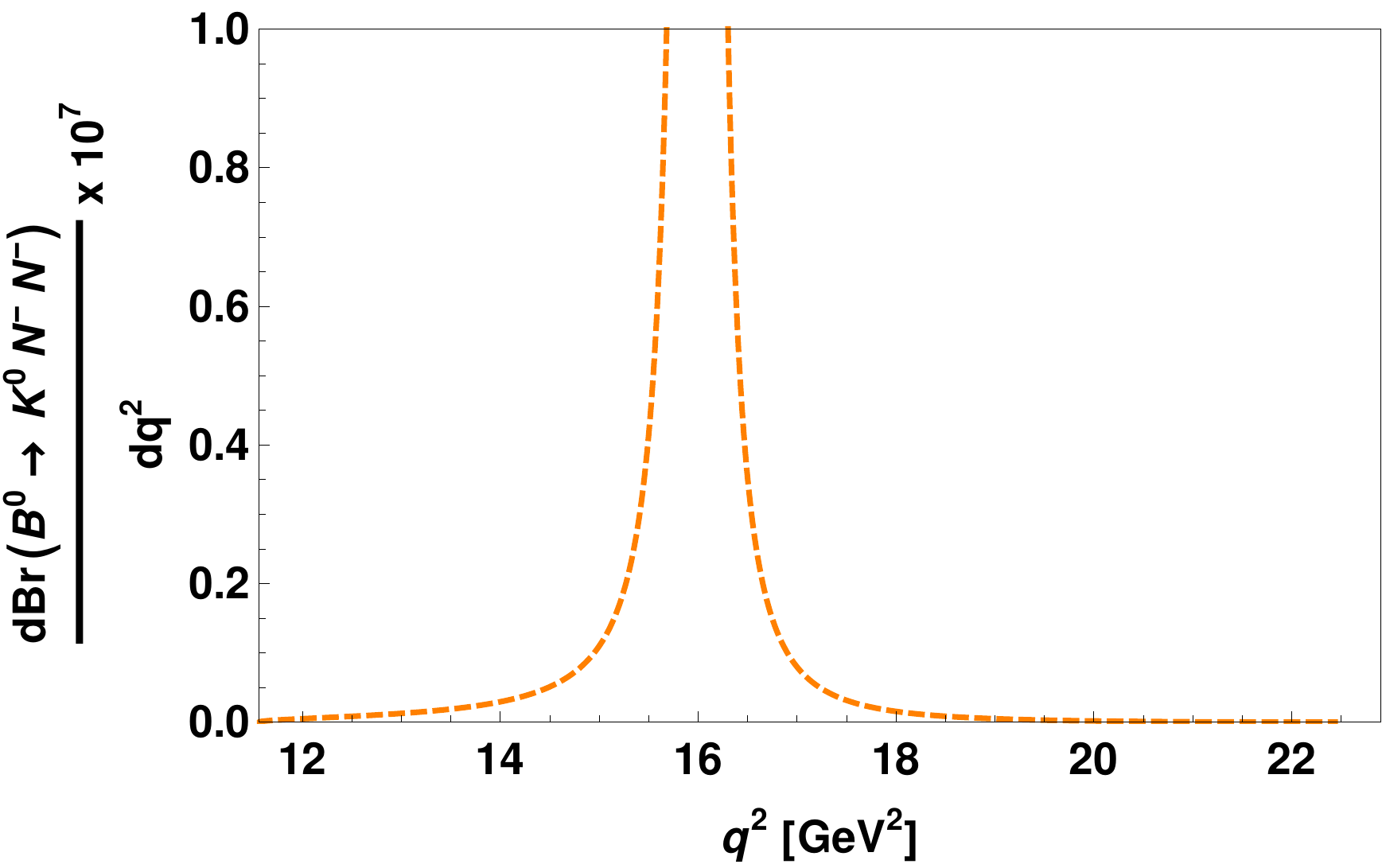}
\quad
\includegraphics[width=0.48\linewidth]{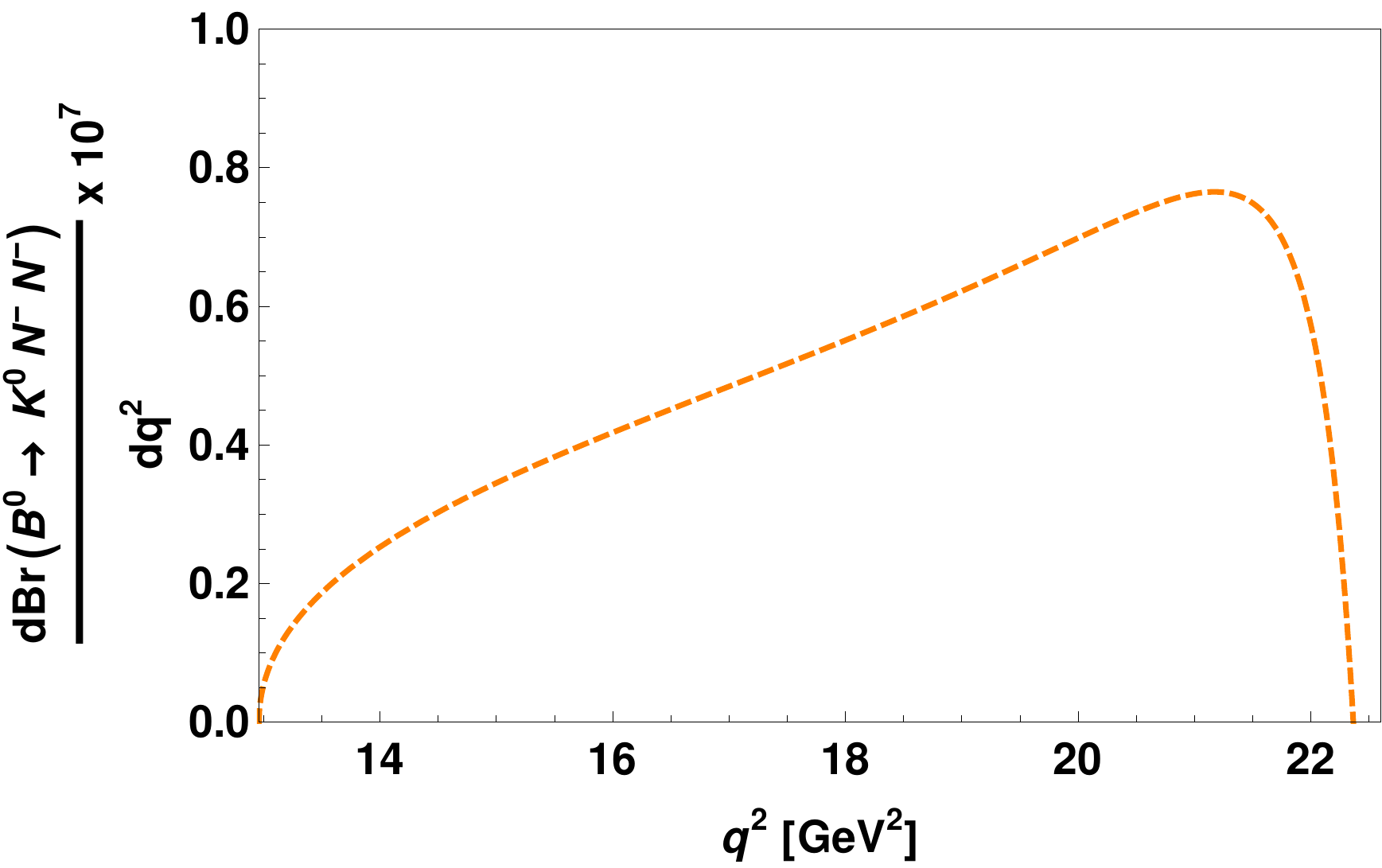}
\quad
\includegraphics[width=0.48\linewidth]{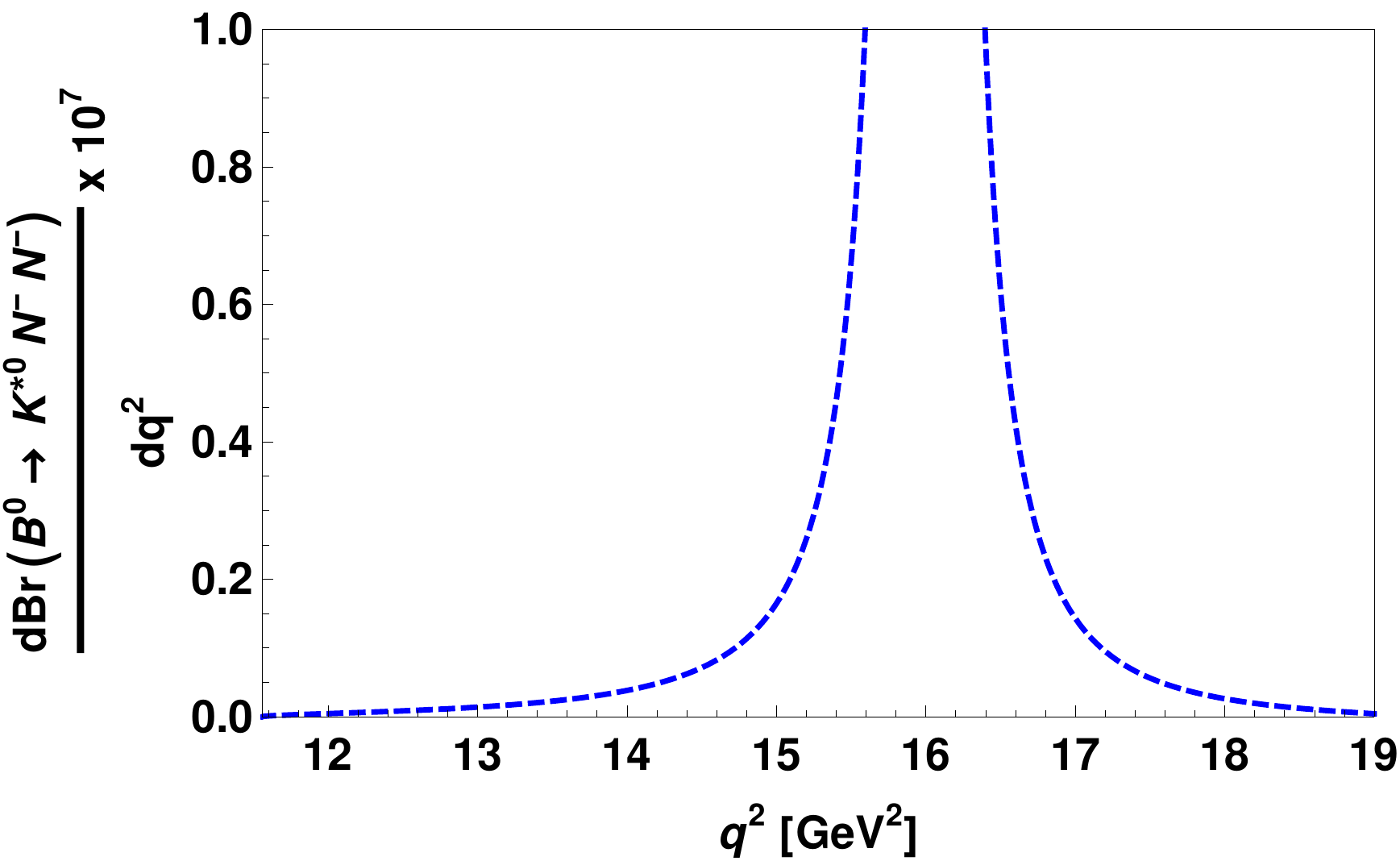}
\quad
\includegraphics[width=0.48\linewidth]{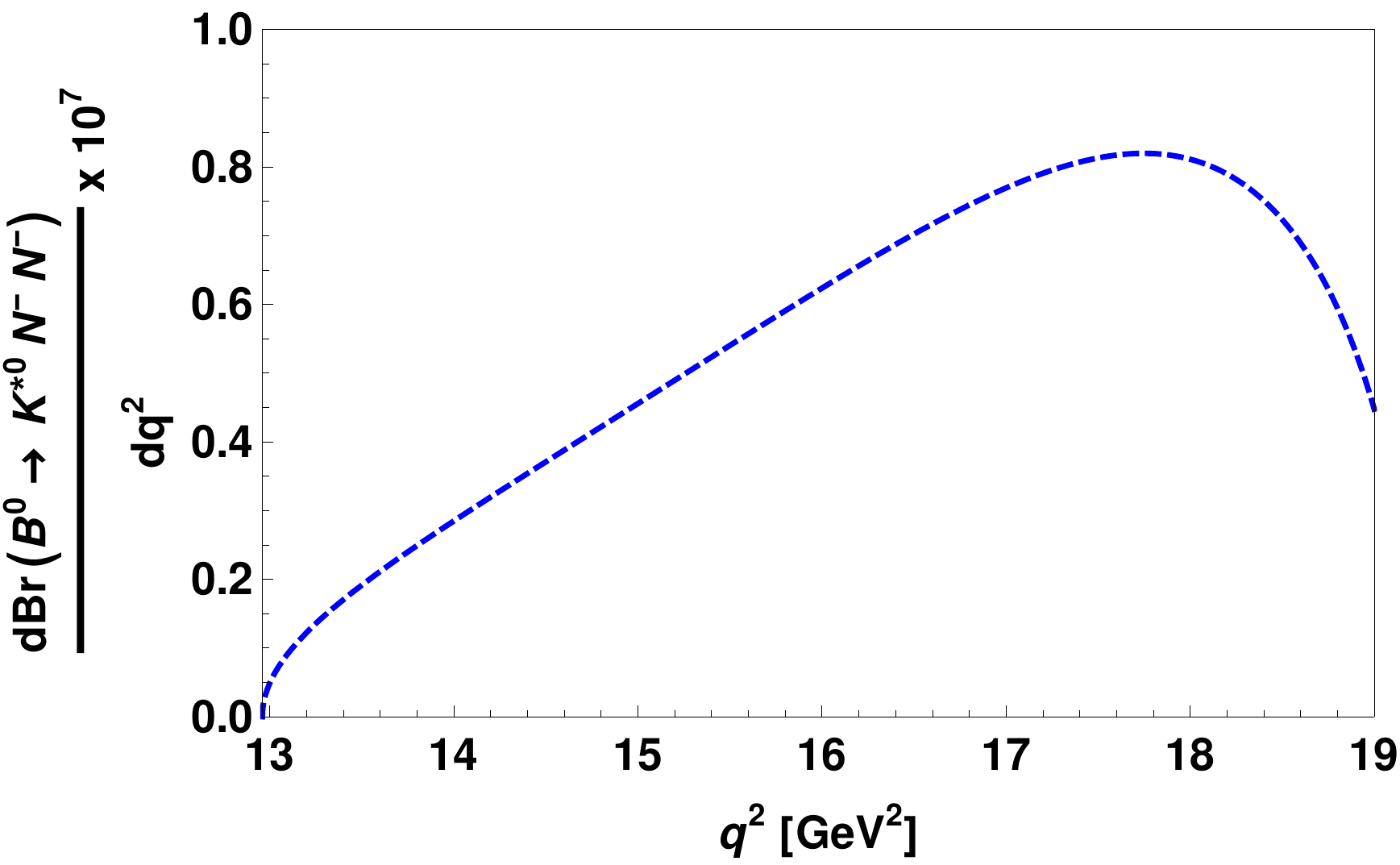}
\quad
\includegraphics[width=0.48\linewidth]{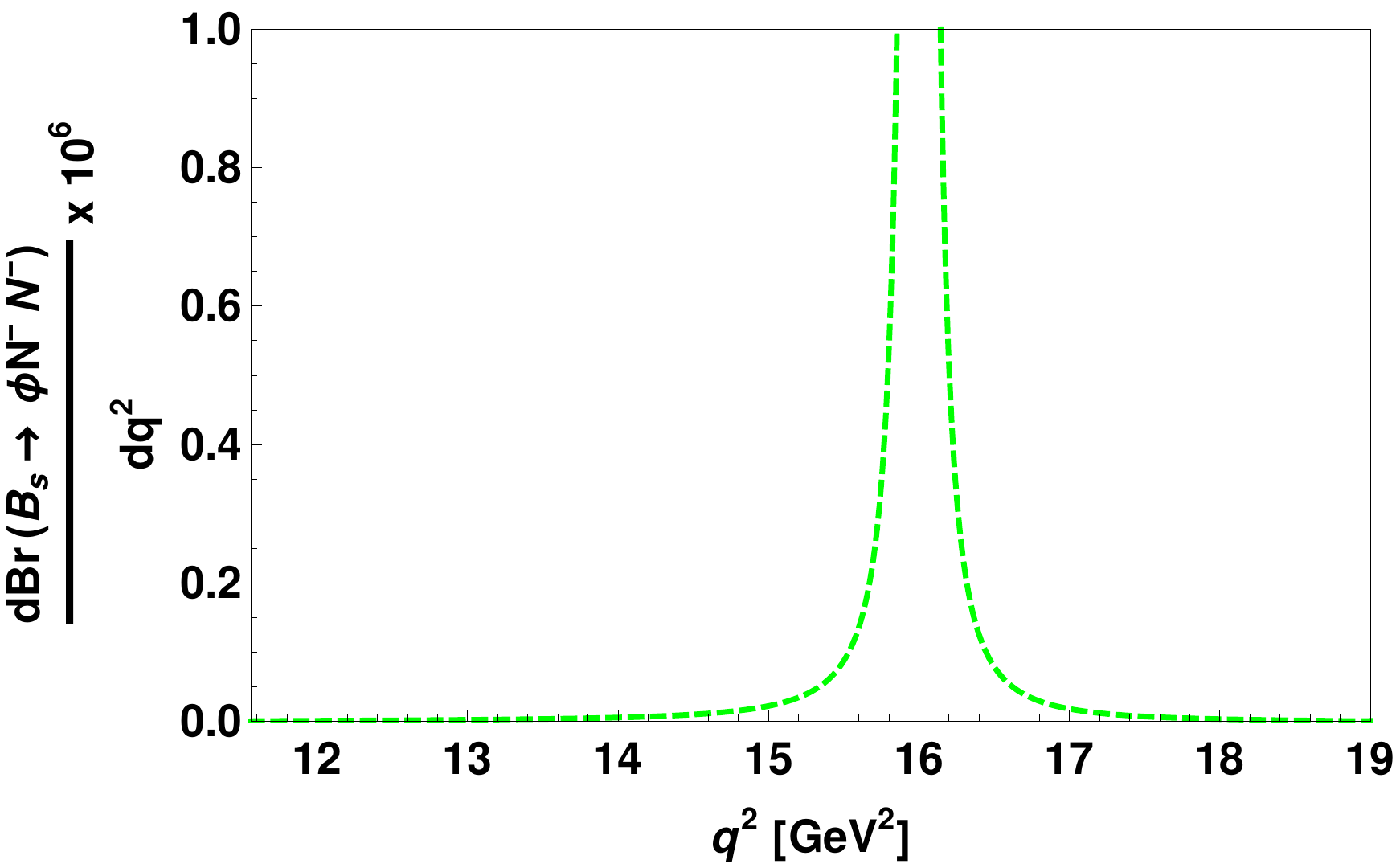}
\quad
\includegraphics[width=0.48\linewidth]{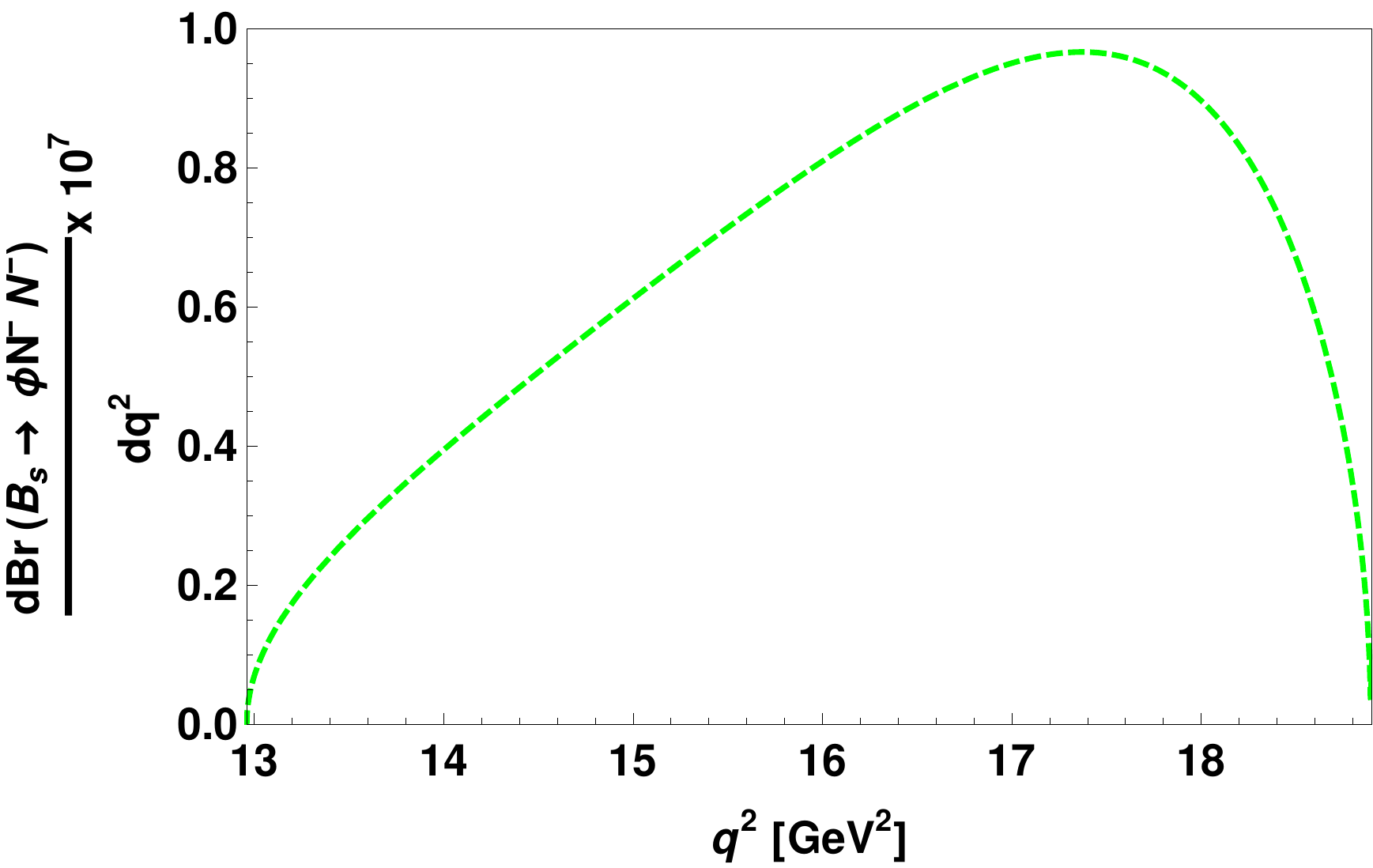}
\caption{The branching ratios of $B^0 \to K^0 N_-N_-$ (left-top panel), $B^0 \to K^{*0} N_-N_-$ (left-middle panel) and $B_s \to \phi N_-N_-$ (left-bottom panel) with respect to $q^2$ for Benchmark-I.  The corresponding plots obtained by using the Benchmark-II are presented in the right panel.  }
\label{Fig:result}
\end{center}
\end{figure}
By using all the discussed input parameters, we show the branching ratios of $B^0 \to K^0 N_-N_-$ (top panel), $B^0 \to K^{*0} N_-N_-$ (middle panel) and $B_s \to \phi N_-N_-$ (bottom panel) with respect to $q^2$ in Fig. \ref{Fig:result}\,. Here, the left panel figures are obtained by using the Benchmark-I values and the right panel plots are for  Benchmark-II values. The estimated numerical values of the branching ratios of $b \to s N_-N_-$ processes for both sets of benchmark values are tabulated in Table \ref{Tab:result}\,. For Benchmark-I, there is a singularity at $q^2=M_{Z^\prime}^2$, i.e., $q^2=16~{\rm GeV}^2$. In order to avoid the singularity, we put the cuts at $(M_{Z'}-0.002)^2\leq q^2 \leq (M_{Z'}+0.002)^2 $ ${\rm GeV}^2$.
\begin{table}[htb]
\centering
\begin{tabular}{|c|c|c|c|c|c|c|}
\hline
Br($b \to s N_- N_-$)~&~ Values using Benchmark-I~&~Values using Benchmark-II ~\\
\hline \hline
Br($B^0 \to K^0 N_-N_-$)~&~$1.92\times 10^{-6}$~&~$4.1\times 10^{-7}$~~\\
\hline
Br($B^+ \to K^+ N_-N_-$)~&~$2.072 \times 10^{-6}$~&~$2.6\times 10^{-7}$~\\
\hline
Br($B^0 \to K^{* 0} N_-N_-$)~&~$3.23\times 10^{-6}$~&~$3.3\times 10^{-7}$~~\\
\hline
Br($B^+\to K^{* +} N_-N_-$)~&~$3.51\times 10^{-6}$~&~$3.61\times 10^{-7}$~\\
\hline
Br($B_s \to \phi N_-N_-$)~&~$4.18\times 10^{-6}$~&~$3.93\times 10^{-7}$~~\\
\hline
\end{tabular}
\caption{The predicted branching ratios of $b \to s N_-N_-$ processes for two different benchmark values of new parameters, which are compatible with both the dark matter and the flavor sectors. }
\label{Tab:result}
\end{table}
In Table \ref{Tab:result-ME}\,, we present the branching ratios of $b \to s \displaystyle{\not} E$ which are the sum of  the branching ratios of $b \to s \nu_l \bar \nu_l$ and $b \to s N_-N_-$ decay processes. We observe that, the addition of $b \to s N_-N_-$ process provide deviation from the SM predictions and are within the experimental limits. 
\begin{table}[htb]
\centering
\begin{tabular}{|c|c|c|c|c|c|c|}
\hline
Br($b \to s \displaystyle{\not} E$)~&~Benchmark-I~&~Benchmark-II~&~Experimental Limit \cite{Zyla:2020zbs}\\
\hline \hline
Br($B^0 \to K^0 \displaystyle{\not} E$)~&~$0.645\times 10^{-5}$~&~$0.457\times 10^{-5}$~&~$<2.6\times 10^{-5}$~\\
\hline
Br($B^+ \to K^+ \displaystyle{\not} E$)~&~$0.697\times 10^{-5}$~&~$0.516\times 10^{-5}$~&~$<1.6\times 10^{-5}$\\
\hline
Br($B^0 \to K^{* 0} \displaystyle{\not} E$)~&~$1.271\times 10^{-5}$~&~$0.981\times 10^{-5}$~&~$<1.8\times 10^{-5}$~\\
\hline
Br($B^+\to K^{* +} \displaystyle{\not} E$)~&~$1.381\times 10^{-5}$~&~$1.066\times 10^{-5}$~&~$<4.0\times 10^{-5}$~\\
\hline
Br($B_s \to \phi \displaystyle{\not} E$)~&~$1.618\times 10^{-5}$~&~$1.24\times 10^{-5}$~&~$<5.4\times 10^{-3}$~\\
\hline
\end{tabular}
\caption{The predicted branching ratios of $b \to s \displaystyle{\not} E$ processes for two different benchmark values of new parameters.  }
\label{Tab:result-ME}
\end{table}

\section{Conclusion}
In this work, we have investigated light GeV scale dark matter and flavor anomalies in a simple $U(1)_{L_\mu - L_\tau}$ variant with heavy neutral fermions and a $(\overline{3},1,1/3)$ scalar leptoquark. The $U(1)$ associated gauge boson ($Z^\prime$) plays a key role and is explored in the low mass regime. The lightest fermion is a stable dark matter and the resonance in $Z^\prime$ portal annihilation channels brings down the relic density to meet Planck data. WIMP-nucleon cross section of spin-dependent type is obtained in leptoquark portal and is looked up for consistency with CDMSlite bound. A benchmark is provided for generating light neutrino mass radiatively with small Yukawa. 

We have constrained the new parameters by using the the branching ratios of $b \to sll$, $b \to s \gamma$ and the $R_{K^{(*)}}$ observables. We have taken two different sets of benchmark values of new parameters (which are compatible with both the dark matter and flavor phenomenology) and have shown the impact on rare $B$ meson decays to missing energy. There exist only experimental upper limits on the branching ratios of $b \to s +$missing energy processes. In the SM, the missing energy  can be carried away by a pair of neutrino, i.e. by $b \to s \nu_l \bar \nu_l$ processes.  We have assumed  the missing energy part as a pair of dark matter in the proposed ${L_\mu - L_\tau}$ scenario. We have shown our prediction for the branching ratios of $b \to s \displaystyle{\not} E$ for two sets of benchmark values which are within the experimental limits. The observation of these modes at LHCb and Belle-II experiments would provide strong hints for the existence of light fermionic dark matter.

\acknowledgments 
S. Singirala and RM would like  to thank University of Hyderabad for financial support through the IoE project grant IoE/RC1/RC1-20-012. RM acknowledges the support from SERB, Govt. of India through grant No, EMR/2017/001448. 
\appendix
\section{Relevant vertices and couplings}
\begin{table}[h]
\begin{center}
\begin{tabular}{|c|c|c|c|c|}
	\hline
			 Vertex	& Coupling \\
\hline
	 $\overline{\mu} \, \gamma^\mu (c^\mu_V - c^\mu_A\gamma^5) \mu Z_\mu$ & $\frac{g}{2\cos\theta_w}(\cos\alpha - \sin\alpha \sin\theta_w\tan\chi)$ \\
	 $\overline{\tau} \, \gamma^\mu (c^\tau_V - c^\tau_A\gamma^5) \tau Z_\mu$ & $\frac{g}{2\cos\theta_w}(\cos\alpha - \sin\alpha \sin\theta_w\tan\chi)$ \\
  $ \overline{\nu_{\mu}} \, \gamma^\mu (c^{\nu_\mu}_V - c^{\nu_\mu}_A\gamma^5) \nu_{\mu} Z_\mu$ & $\frac{g}{2\cos\theta_w}(\cos\alpha - \sin\alpha \sin\theta_w\tan\chi)$ \\
   $ \overline{\nu_{\tau}} \, \gamma^\mu (c^{\nu_\tau}_V - c^{\nu_\tau}_A\gamma^5) \nu_{\tau} Z_\mu$ & $\frac{g}{2\cos\theta_w}(\cos\alpha - \sin\alpha \sin\theta_w\tan\chi)$ \\
   $\overline{N_-} \gamma^\mu \gamma^5 N_{-} Z_\mu$ & $ g_{\mu\tau} \cos 2\beta \sin\alpha \sec\chi$ \\
   $\overline{N_+} \gamma^\mu \gamma^5 N_{+} Z_\mu$ & $- g_{\mu\tau}\cos 2\beta \sin\alpha \sec\chi $ \\
      $\left(\overline{N_-} \gamma^\mu \gamma^5 N_{+} + \overline{N_+} \gamma^\mu \gamma^5 N_{-}\right) Z_\mu$ & $ g_{\mu\tau}\sin 2\beta \sin\alpha \sec\chi $ \\
   $\left(S_1\d^\mu S_1^\dagger - S_1^\dagger \d^\mu S_1 \right) Z_\mu$ & $\frac{g^\prime}{3}(-\cos\alpha \sin\theta_w+\sin\alpha\tan\chi) + g_{\mu\tau} \sin\alpha \sec\chi$ \\
   \hline
	 $\overline{\mu} \, \gamma^\mu \mu A_\mu$ & $g \sin\theta_w$  \\
	 $\overline{\tau} \, \gamma^\mu \tau A_\mu$ & $g \sin\theta_w$  \\
   $\left(S_1\d^\mu S_1^\dagger - S_1^\dagger \d^\mu S_1 \right) A_\mu$ & $\frac{g^\prime}{3}\cos\theta_w$ \\
	\hline
	 $\overline{\mu} \, \gamma^\mu \mu Z^\prime_\mu$ & $g_{\mu\tau} \cos\alpha \sec\chi$  \\
	 $\overline{\tau} \, \gamma^\mu \tau Z^\prime_\mu$ & -$g_{\mu\tau}\cos\alpha \sec\chi$  \\
  $ \overline{\nu_{\mu}} \, \gamma^\mu (1-\gamma^5) \nu_{\mu} Z^\prime_\mu$ & $g_{\mu\tau}\cos\alpha \sec\chi$  \\
   $ \overline{\nu_{\tau}} \, \gamma^\mu (1-\gamma^5) \nu_{\tau} Z^\prime_\mu$ & -$g_{\mu\tau}\cos\alpha \sec\chi$  \\
   $\overline{N_-} \gamma^\mu \gamma^5 N_{-} Z^\prime_\mu$ & -$ g_{\mu\tau} \cos 2\beta \cos\alpha \sec\chi$  \\
   $\overline{N_+} \gamma^\mu \gamma^5 N_{+} Z^\prime_\mu$ & $ g_{\mu\tau}\cos 2\beta \cos\alpha \sec\chi $  \\
      $\left(\overline{N_-} \gamma^\mu \gamma^5 N_{+}  + \overline{N_+} \gamma^\mu \gamma^5 N_{-} \right)Z^\prime_\mu$ & -$ g_{\mu\tau}\sin 2\beta \cos\alpha \sec\chi $  \\
   $\left(S_1\d^\mu S_1^\dagger - S_1^\dagger \d^\mu S_1 \right) Z^\prime_\mu$ & $-\frac{g^\prime}{3}(\sin\alpha \sin\theta_w + \cos\alpha\tan\chi) - g_{\mu\tau} \cos\alpha \sec\chi$ \\
   \hline
   \hline
\end{tabular}
\caption{Fermion-gauge vertices and couplings.}
\label{Gauge_vertices}
\end{center}
\end{table}
\begin{table}[h]
\begin{center}
\begin{tabular}{|c|c|c|}
	\hline
			 Vertex	& Coupling  \\
\hline
	 $\overline{\mu} \mu H_1$ & $\frac{m_\mu}{v} \cos\zeta$  \\
	 $\overline{\tau}\tau H_1$ & $\frac{m_\tau}{v} \cos\zeta$  \\
      $\overline{N^c_-} N_{-} H_1$ & $-\frac{1}{\sqrt{2}}\left(f_\mu \cos^2 \beta+ f_\tau \sin^2\beta\right)\sin\zeta$  \\
      $\overline{N^c_+} N_{+} H_1$ & $-\frac{1}{\sqrt{2}}\left(f_\mu \sin^2 \beta+ f_\tau \cos^2\beta\right)\sin\zeta$  \\
         $\overline{N^c_-} N_{+} H_1$ & $-\frac{1}{\sqrt{2}}\left(f_\mu- f_\tau\right)\sin2\beta\sin\zeta$  \\
         $S_1^\dagger S_1 H_1$ & $\lambda_{HS}v \cos\zeta - \lambda_{S\phi}v_2 \sin\zeta$  \\
      \hline
	 $\overline{\mu} \mu H_2$ & $\frac{m_\mu}{v} \sin\zeta$  \\
	 $\overline{\tau}\tau H_2$ & $\frac{m_\tau}{v} \sin\zeta$  \\
      $\overline{N^c_-} N_{-} H_2$ & $\frac{1}{\sqrt{2}}\left(f_\mu \cos^2 \beta+ f_\tau \sin^2\beta\right)\cos\zeta$  \\
      $\overline{N^c_+} N_{+} H_2$ & $\frac{1}{\sqrt{2}}\left(f_\mu \sin^2 \beta+ f_\tau \cos^2\beta\right)\cos\zeta$  \\
         $\overline{N^c_-} N_{+}H_2$ & $\frac{1}{\sqrt{2}}\left(f_\mu- f_\tau\right)\sin2\beta\cos\zeta$  \\
         $S_1^\dagger S_1 H_2$ & $\lambda_{HS}v \sin\zeta + \lambda_{S\phi}v_2 \cos\zeta$  \\
	\hline
	$\ol{d_{qR}^c} S_1 N_{-} + {\rm{h.c.}}$ & $y_{qR}\cos\beta$\\ 
	$\ol{d_{qR}^c} S_1 N_{+} + {\rm{h.c.}}$ & $y_{qR}\sin\beta$\\ 
	\hline
	\hline
\end{tabular}
\caption{Fermion-scalar vertices and couplings.}
\label{scalar_vertices}
\end{center}
\end{table}

\section{Loop functions}
The $b \to sll$ loop function, used in Section VI is given by
\bea
 \mathcal{V}_{sb}(\chi_-, \chi_+)&=&\sin^22\beta \cos\alpha \sec\chi \left(1+4\sqrt{\chi_-\chi_+}j\left(\chi_-,\chi_+\right)-2k\left(\chi_-,\chi_+\right)\right)\nn \\&+&2(\sin^2\beta I(\chi_+)+\cos^2\beta  I(\chi_-))\cos^2\beta \cos\alpha \sec\chi\,,
\eea
where 
\bea \label{A:loop-2}
f(\chi_1, \chi_2, \chi_3,\cdots)\equiv\frac{f(\chi_1, \chi_3,\cdots)-f(\chi_2, \chi_3,\cdots)}{\chi_1-\chi_2},~~~~~~~~f=j,\kappa\,,
\eea
with
\bea \label{A:loop-3}
j(\chi)&=&\frac{\chi \log \chi}{\chi-1}\,,\\\label{A:loop-4}
\kappa(\chi)&=&\frac{\chi^2 \log \chi}{\chi-1}\,,\\\label{A:loop-5}
I(\chi)&=&\frac{-3\chi^2+4\chi-1+2\chi^2\log \chi}{8(\chi-1)^2}\,.
\eea 

\section{Effective Wilson coefficients}
The effective $C_7^{\rm eff}$ and $C_9^{\rm eff}$ Wilson coefficients including the four-quark and gluon dipole operators, as mention in Eqn. (25) are given as \cite{Bobeth:2011gi}
\bea
C_7^{\rm eff}&=&C_7-\frac{1}{3} \left[C_3+\frac{4}{3}C_4+20 C_5+\frac{80}{3} C_6 \right]+ 
\frac{\alpha_s}{4\pi}\left[\left(C_1-6 C_2\right)A(q^2)-C_8 F_8^{(7)}(q^2)\right],~\\
C_9^{\rm eff}&=&C_9+h\left(0, q^2\right) \left[\frac{4}{3}C_1+C_2+\frac{11}{2}C_3 - \frac{2}{3}C_4 +52C_5 -\frac{32}{3}C_6\right]\nn\\
&-&\frac{1}{2}h\left(m_b, q^2\right)\left[7C_3+\frac{4}{3}C_4+76C_5+\frac{64}{3}C_6\right]+ \frac{4}{3}\left[C_3+\frac{16}{3}C_5
+\frac{16}{9}C_6\right]\nn\\
&+&\frac{\alpha_s}{4\pi}\left[C_1\left(B(q^2)+4C(q^2)\right)-3C_2\left(2B\left(q^2\right)-C\left(q^2\right)\right)-C_8F_8^{(9)}\left(q^2\right)\right]
\nn\\
& +& 8\frac{m^2_c}{q^2}\left[\left(\frac{4}{9}C_1+\frac{1}{3}C_2\right)(1+\lambda_u) +2C_3+20C_5\right]\;,
\eea
where $\lambda_u = ({V_{ub}V^*_{us}})/{(V_{tb}V^*_{ts})}$ and the functions $h(m_i, q^2)$ and $A, B, C,F_8^{(7,9)}$ can be found in Ref. \cite{Grinstein:2004vb, Bobeth:2010wg}.

\bibliography{BL}

\end{document}